\documentclass{article}
\usepackage{amsmath,amssymb}
\usepackage{graphicx}%
\newtheorem{thm}{Statement}
\newcommand{\Fig}[3]{%
\begin{center}
\parbox{#2cm}{%
\refstepcounter{figure}\includegraphics[width=#2cm]{#1}\\[12pt] \noindent {\small Figure \thefigure:\quad
#3}}\end{center}}
\newcommand{\FigReg}[4]{%
\begin{center}
\parbox{#2cm}{%
\refstepcounter{figure}\includegraphics[width=#2cm,height=#3cm]{#1} \\[12pt]\noindent {\small Figure \thefigure:\quad
#4}}\end{center}}

\begin{document}
\begin{center}
{\bf \Large Qualitative Analysis and Numerical Simulation of Equations of the Standard Cosmological Model} \\[12pt]
Yurii Ignat'ev\\
N.I. Lobachevsky Institute of Mathematics and Mechanics, Kazan Federal University, \\ Kremleovskaya str., 35, Kazan, 420008, Russia
\end{center}

\begin{abstract}
On the basis of qualitative theory of differential equations it is shown that dynamic system based on the system of Einstein - Klein - Gordon equations with regard to Friedman Universe has a stable center corresponding to zero values of scalar potential and its derivative at infinity. Thus, the cosmological model based on single massive classical scalar field in infinite future would give a flat Universe. The carried out numerical simulation of the dynamic system corresponding to the system of Einstein - Klein - Gordon equations showed that at great times of the evolution the invariant cosmological acceleration has a microscopic oscillating character ($T\sim 2\pi mt$), while macroscopic value of the cosmological acceleration varies from $+1$ at inflation stage after which if decreases fast to $-1/2$ (non-relativistic stage), and then slowly tends to $-1$ (ultrarelativistic stage).

{\bf keyword} standard cosmological model, instability, quality analysis, numerical simulation, zero-center flat universe
\end{abstract}

\section{Basic Relations of the Cosmological Model}
In this section we reproduce basic relations and concepts of the standard cosmological model(SCM), which will be required further.
\subsection{Basic Relations for the Space-Flat Friedmann Model}
\hfill A. Einstein in 1917 modified his gravitation theory \cite{Einstein17}, having added the so-called cosmological term ($\Lambda$ - term) to the equations of the gravitational field\footnote{Here and further the Planck system of units $\hbar=c=G=1$ is chosen; signature metric is $(-1,-1,-1,+1)$, Ricci tensor is obtained from Riemann tensor by means of the first and the third indices' convolution.}:
\begin{equation}\label{EinstEqL}
G^i_k=\Lambda\delta^i_k+8\pi T^i_k.
\end{equation}
In the same year de Sitter found the solution of these equations for the constant curvature space \cite{deSitter17}:
\begin{equation}
\label{DeSitter}
ds^2=\cos^2\frac{r}{R}dt^2-dr^2-R^2\sin^2\frac{r}{R}(d\theta^2+\sin^2\theta d\varphi^2).
\end{equation}
 In 1918 Einstein stated that de Sitter solution (\ref{DeSitter}) has a true singularity at $r = \pi/2 R$ \cite{Einstein18}. This circumstance further had led him to reject of $\Lambda$ - term.

Further it turned out that de Sitter solution's singularity was caused by unfortunate choice of reference frame where light cone was not covering the entire space. In the synchronous frame of reference metrics (\ref{DeSitter}) can be re-written as Friedmann metric \cite{Friedmann1922}. The corresponding solution is called inflation and describes exponentially fast expansion of the Universe:
\begin{equation}
\label{Friedman}
ds^2=dt^2-a^2(t)[dr^2+\varrho^2(r)(d\theta^2+\sin^2\theta d\varphi^2)];
\end{equation}
\begin{equation}
\label{a_rho}
\begin{array}{llll}%
a(t)=a_0 H_0^{-1}\cosh H_0t;& \varrho(r)=\sin r;& k>0;\\[6pt]
a(t)=a_0 \mathrm{e}^{H_0t};& \varrho(r)=r;& k=0;\\[6pt]
a(t)=a_0H_0^{-1}\sinh H_0t;& \varrho(r)=\mathrm{sh} r;& k<0,\\
\end{array}
\end{equation}
where $k$ is a curvature of the 3-dimensional space and
\begin{equation}\label{H_0}
H_0=\sqrt{\Lambda/3}.
\end{equation}
Metrics (\ref{Friedman}) can be re-written in form of metrics conformally corresponding to the space of constant curvature:
\begin{equation}
\label{a(eta)}
ds^2=a^2(\eta)[d\eta^2-dr^2+\varrho^2(r)(d\theta^2+\sin^2\theta d\varphi^2)],
\end{equation}
where
\begin{equation}
\label{eta(t)}
a(\eta)d\eta=dt \Leftrightarrow t=\int a(\eta)d\eta \Leftrightarrow \eta=\int\frac{dt}{a(t)}.
\end{equation}
In the case of zero space curvature $\varrho=r$ the Friedmann metrics (\ref{Friedman}) takes form
\begin{equation}\label{Friedman0}
ds^2=dt^2-a^2(t)(dx^2+dy^2+dz^2)
\end{equation}
and the unique non-trivial Einstein equation \emph{for the flat Universe} takes the following form:
\begin{equation}
\label{Einst_0}
\frac{a'^2}{a^4}=\Lambda\Rightarrow \frac{a'}{a^2}=\sqrt{\Lambda} .
\end{equation}
The solution of this equation is:
\begin{equation}\label{a(t)_Lambda}
a=\frac{a_0}{1-a_0\sqrt{\Lambda}(\eta-\eta_0)}
\end{equation}
Choosing constant $\eta_0$ so that singular (zero) value of the scale factor corresponds to $\eta=-\infty$
(i.e., $\eta_0a_0\sqrt{\Lambda}=1$), we find
\begin{equation}
\label{sol_a(eta)}
a(\eta)=-\frac{1}{\sqrt{\Lambda}\eta},\quad \eta\in (-\infty,-0).
\end{equation}
Using (\ref{eta(t)}) let us find the relation between time variable $\eta$ and physical time $t$.
Infinitely far past corresponds to $\eta=-\infty$ and infinitely far future corresponds to $\eta=-0$.
Integrating equation (\ref{a(eta)}) with a help of solution (\ref{sol_a(eta)}), we obtain
\begin{eqnarray}
\label{t(eta)}
t=-\frac{1}{\sqrt{\Lambda}}\ln(-\eta), \Rightarrow \eta\to-\infty\longleftrightarrow \nonumber\\
t \to-\infty; \; \eta\to+\infty \longleftrightarrow t\to+\infty.
\end{eqnarray}
Thus, the history of the de Sitter Universe does not have neither beginning nor the end.
Substitution of (\ref{t(eta)}) into (\ref{sol_a(eta)}) exactly leads to inflation solution (\ref{a_rho}) for the space-flat Universe:
\begin{equation}\label{Lambda-sol}
a(t)=a_0 \mathrm{e}^{H_0t}.
\end{equation}

Further we agree the following standard notations for derivative functions over time variables:
\begin{equation}\label{dot'}
f'=\frac{\partial f}{\partial\eta},\; \dot{f}=\frac{\partial f}{\partial t}\Rightarrow f'=a\dot{f};\; \dot{f}=\frac{f'}{a}.
\end{equation}
\subsection{The Energy-Momentum Tensor}
Let us take into account a contribution of dark matter in the right side of the Einstein equations (\ref{EinstEqL}).
Metric's symmetries with respect to rotations and transpositions of 3-dimensionless space as a result of the Einstein equations lead to identical symmetries of the energy-momentum tensor. Therefore the energy-momentum tensor of the homogenous isotropic Universe must have the structure of ideal flux'es energy-momentum tensor:
\begin{equation}\label{Tik}
T^i_k=(\varepsilon+p)\delta^i_4\delta^4_k-p\delta^i_k.
\end{equation}
In accordance with (\ref{Tik}) E.B.Gliner \cite{Gliner} interpreted the cosmological term as energy density of vacuum with the following equation of state:
\begin{equation}\label{lambda-e}
\Lambda=\varepsilon_0 \Rightarrow \varepsilon_0+p_0=0.
\end{equation}
As a result there remain two independent Einstein equations one of which in consequence of Bianchi identities can be replaced with the differential implication - the energy conservation law:
\begin{equation}\label{e'}
\varepsilon' + 3\frac{a'}{a}(\varepsilon+p)=0\Leftrightarrow \dot{\varepsilon} + 3\frac{\dot{a}}{a}(\varepsilon+p)=0.
\end{equation}
At given \emph{equation of state}
\begin{equation}\label{eq_state}
p=p(\varepsilon)
\end{equation}
equation (\ref{e'}) is always integrated in quadratures:
\begin{equation}\label{deda}
\int\frac{d\varepsilon}{\varepsilon+p(\varepsilon)}= -3\ln a.
\end{equation}
In particular, at linear relation between pressure and energy density ($\kappa$ is called a barotropic coefficient)
\begin{equation}\label{barotrop}
p=\kappa\varepsilon
\end{equation}
the Einstein equations are integrated in elementary functions. In presence of matter the remaining Einstein equation takes the form:
\begin{equation}
\label{Einst_E}
3\frac{a'^2}{a^4}=\Lambda+8\pi\varepsilon\Rightarrow 3\frac{\dot{a}^2}{a^2}=\Lambda+8\pi\varepsilon.
\end{equation}
{\it Notice}: The system of equations (\ref{e'}), (\ref{Einst_E}) is invariant with respect to scale transformation
\begin{equation}\label{a->a}
a(t)\to\mathrm{ Const}\cdot a(t),
\end{equation}
therefore scale function $a(t)$ can be assumed equal to, for instance, $1$ at arbitrary non-singular instant of physical time $t$.

Let us also introduce two scalar functions in terms of which it is convenient to analyze the cosmological evolution:
$H(t)$ is a Hubble constant
\begin{equation}\label{H}
H=\frac{\dot{a}}{a}=\frac{a'}{a^2},
\end{equation}
and invariant cosmological acceleration:
\begin{equation}\label{Omega}
\Omega=\frac{\ddot{a}a}{\dot{a}^2}\equiv 1+\frac{\dot{H}}{H^2}
\end{equation}
Let us also notice the important relation between $\Omega$ and barotropic coefficient $\kappa$:
\begin{equation}\label{Omega-kappa}
\Omega = -\frac{1}{2} (1+3\kappa).
\end{equation}
\section{The Analysis of the Standard Model with a Scalar Field}
\subsection{The Equations of the Cosmological Model with a Scalar Vacuum}
In the standard cosmological model of the early Universe (SCM) the classical massive scalar field $\Phi$ corresponded by following energy-momentum tensor is considered as a model of vacuum:
\begin{equation}
\label{T_{iks}}
T^i_k=2\Phi^{,i}\Phi_{,k}-\delta^i_k\Phi_{,j}\Phi^{,j}+\delta^i_k m^2\Phi^2,
\end{equation}
where $m$ is a mass of this field's quanta. The equation of the scalar field is obtained from Bianchi identity laws  $T^i_{k,i}=0$:
\begin{equation}\label{d2F}
\Box\Phi+m^2\Phi=0,
\end{equation}
where
\begin{equation}
\label{dAlamber}
\Box\Phi=g^{ik}\Phi_{,ik}\equiv \frac{1}{\sqrt{-g}}\frac{\partial}{\partial x^i}\sqrt{-g}g^{ik}\frac{\partial}{\partial x^k}\Phi
\end{equation}
is D'Alembert operator. For the homogenous Universe it is $\Phi=\Phi(t)$, and the scalar field's energy-momentum takes isotropic structure
(\ref{Tik}), where
\begin{equation}
\label{eps}
\varepsilon=\dot{\Phi}^2+m^2\Phi^2; \quad p=\dot{\Phi}^2-m^2\Phi^2,
\end{equation}
so that
\begin{equation}\label{e+p}
\varepsilon+p=2\dot{\Phi}^2,
\end{equation}
and field equation (\ref{d2F}) takes the form:
\begin{equation}
\label{d2F(t)}
\ddot{\Phi}+3\frac{\dot{a}}{a}\dot{\Phi}+m^2\Phi=0.
\end{equation}
At the same time the unique independent Einstein equation is\footnote{Further in the article we consider a model with $\Lambda=0$.}:
\begin{equation}\label{Einst_Eq_S}
3\frac{\dot{a}^2}{a^2}=\Lambda+8\pi(\dot{\Phi}^2+m^2\Phi^2).
\end{equation}
\emph{Note}:\\
Let us notice that if it was $\dot{\Phi}=0$, then according to (\ref{eps}), we would get a vacuum equation of state $\varepsilon+p=0$.
In this case the scalar field could play a role of the effective cosmological term.

\subsection{Approximation of Slow Rolling}
Standard model is based on so-called \emph{approximation of slow rolling} in the equations  (\ref{d2F(t)}) and (\ref{Einst_Eq_S}) (see e.g. \cite{Astar})\footnote{Here we write out basic relations of the standard cosmological model for a case of simplest potential $V(\Phi)=m^2\Phi^2$}:
\begin{equation}\label{slow_roll}
|\dot{H}|\ll H^2\Longleftrightarrow |\Omega-1|\ll 1 \Longleftrightarrow \Omega\to1,
\end{equation}
Thus, the approximation of slow rolling is equivalent to impose of the condition on the second derivatives of the scale factor:
\begin{equation}\label{slow_eq}
a\ddot{a}=\dot{a}^2.
\end{equation}
The relation (\ref{slow_eq}) is an autonomous differential equation on the scale factor. Carrying out simplest transformation:
\begin{equation}\label{t->a}
\ddot{a}\equiv \frac{d\dot{a}}{da}\dot{a},
\end{equation}
let us reduce it to the form:
\[\frac{d\dot{a}}{\dot{a}}=\frac{da}{a}\]
and find its solution:
\[\dot{a}=\mathrm{C}a,\]
where $\mathrm{C}$ is an arbitrary constant. Integrating this relation one more time we find the inflation solution for the scale factor:
\begin{equation}\label{inflat_K}
a(t)=a_1\mathrm{e}^{\mathrm{C}t}.
\end{equation}
In this case it is:
\begin{equation}\label{K,Omega}
H=H_0=\mathrm{C};\quad \Omega=1.
\end{equation}

\begin{thm}
Thus, the so-called approximation of slow rolling (\ref{slow_roll}) is fully equivalent to suggestion about that the Einstein equations' solution is asymptotically inflation one:
\begin{equation}\label{slow_inflat}
\ddot{H}\ll \dot{H}^2 \Leftrightarrow a(t)\varpropto \displaystyle\mathrm{e}^{\mathrm{H_0}t}.
\end{equation}
\end{thm}

\subsection{The Solution of the System of Equation in Approximation of Slow Rolling}
Now it is necessary to clarify how to obtain inflation solution (\ref{inflat_K}) as an \emph{asymptotic solution} of the Einstein equation (\ref{Einst_Eq_S}).
Direct substitution of (\ref{inflat_K}) into Einstein equation (\ref{Einst_Eq_S}) leads to equality:
\begin{equation}\label{const=e}
3H^2_0=8\pi(\dot{\Phi}^2+m^2\Phi^2).
\end{equation}
Since the left side of the equality (\ref{const=e}) is constant and the right side is a sum of squares of two functions, the following is a general solution of (\ref{const=e}):  %
\begin{equation}\label{F=,dF=}
\Phi=\sqrt{\frac{3}{8\pi}}\frac{H_0}{m}\sin\psi(t);\quad \dot{\Phi}=\sqrt{\frac{3}{8\pi}}\frac{H_0}{m}\cos\psi(t),
\end{equation}
where $\psi(t)$ is so far an arbitrary function. Calculating $\dot{\Phi}$ using differentiation of the first from the relations (\ref{F=,dF=}) and comparing the result with the second one, let us find: $\dot{\psi}=1\rightarrow \psi=t$. Substituting the obtained result into the field equation (\ref{d2F(t)}), we obviously come to relation $H_0/m =0$,
where from we can obtain for the constant $\mathrm{C}$ in the relation (\ref{K,Omega}) the unique possible solution strictly corresponding to the approximation of slow rolling and empty Euclidean space $\mathrm{C}=H_0=0$, $\Phi=0$.
\begin{thm}\label{t:H=0}
Thus, directly applying the approximation of slow rolling we will not achieve necessary result: strictly speaking, the approximation of slow rolling provides only trivial solution of the system of Einstein - Klein - Gordon equations:
\begin{equation}\label{trivial}
\Omega-1\to 0 \leftrightarrow H=0;\quad \Phi=0.
\end{equation}
\end{thm}
\subsection{The Small Mass Approximation}
Therefore in the standard cosmological model to ensure early inflation there is normally used one more approach: the approximation of small mass of the scalar field's quanta:
\begin{equation}\label{mt<<1}
m^2\ll \frac{\dot{a}^2}{a^2}\equiv H^2.
\end{equation}
In this case the following condition should fulfill:
\begin{equation}\label{approx_SM2}
m^2\Phi\ll \ddot{\Phi} \longleftrightarrow m^2\Phi \ll \frac{\dot{a}}{a}\dot{\Phi}.
\end{equation}
In this approximation the equation of the scalar field (\ref{d2F(t)}) takes the following form
\begin{equation}
\label {d2F(t)slow}
\ddot{\Phi}+3\frac{\dot{a}}{a}\dot{\Phi} = 0
\end{equation}
and its \emph{particular solution}, leading to de Sitter vacuum is
\begin{equation}
\label{Sol0}
\Phi=\Phi_0=\mathrm{Const}.
\end{equation}

Herewith according to (\ref{eps}) we obtain:
\begin{equation}\label{p=-e}
\varepsilon=-p=m^2\Phi^2_0=\mathrm{Const}.
\end{equation}
Substituting this solution into Einstein equation (\ref{Einst_Eq_S}) and resolving it we find the inflation solution:
\begin{equation}\label{inflat}
a(t)=a_1 \mathrm{e}^{H_0t},\qquad H_0=\sqrt{\frac{8\pi}{3}}m|\Phi_0|.
\end{equation}

Let us call the solution (\ref{Sol0}) -- (\ref{inflat}) the \emph{standard solution} of Einstein - Klein - Gordon equations in accordance with its leading part in the standard cosmological scenario. In the standard model the beginning of the Universe corresponding to cosmological singularity $a=0$, lays in infiniteäly distant past:
\begin{equation}\label{begin}
a(-\infty)=0.
\end{equation}
Let us notice that the Einstein equation (\ref{Einst_Eq_S}) in the considered model has the form:
\begin{equation}\label{HH0}
H^2=H^2_0,
\end{equation}
therefore according to (\ref{inflat}) condition (\ref{mt<<1}) is equivalent to the condition:
\begin{equation}\label{F>>}
|\Phi_0|\gg \sqrt{\frac{3}{8\pi}},
\end{equation}
i.e. {\it it fulfills only for very big values of the scalar field's potential}.

Let us now carry out next iteration to account massive term in the field equation (\ref{d2F(t)}). To do that let us substitute the inflation solution of the Einstein equation (\ref{inflat}) into the original equation of the scalar field (\ref{d2F(t)})
\begin{equation}\label{Fn1}
\ddot{\Phi}+3H_0\dot{\Phi}+m^2\Phi=0
\end{equation}
and find its exact solution:
\begin{eqnarray}\label{Phi_cor}
\Phi=C_1\exp\biggl[-\frac{1}{2}\bigl(3H_0+\sqrt{9H^2_0-4m^2}\bigr)t\biggr]\nonumber\\
+C_2\exp\biggl[-\frac{1}{2}\bigl(3H_0-\sqrt{9H^2_0-4m^2}\bigr)t\biggr],
\end{eqnarray}
where further we will take into account the solution (\ref{mt<<1}). As a result we obtain:
\begin{equation}\label{Phi1}
\displaystyle\Phi\approx  C_1\mathrm{e}^{- 3H_0t}+C_2 \mathrm{e}^{-\frac{m^2}{3H_0}t}.
\end{equation}
The first term in this solution decays fast at $t\to+\infty$, and the second term with the assumption of small values of time variable
\begin{equation}\label{approx_t}
|m t|\ll \frac{3H_0}{m}\equiv \sqrt{3}|\Phi_0|
\end{equation}
remains approximately constant. Exactly this term ensures the model of the constant scalar field. Therefore in accordance with the more precise model of ``constant'' scalar field in approximation (\ref{mt<<1}) we should suppose:
\begin{equation}\label{Phi1_1}
\Phi\approx \Phi_0 \mathrm{e}^{-\frac{m^2}{3H_0}t}.
\end{equation}
Using the obtained solution instead of standard one (\ref{Sol0}), let us check the fulfillment of the original condition (\ref{approx_SM2}) (condition (\ref{mt<<1}), as we have seen, is reduced to condition (\ref{F>>}) to the value of scalar potential):
\begin{eqnarray}\label{real_cond1}
\ddot{\Phi}\gg m^2\Phi \rightarrow 9H^2_0\ll m^2;\\
\label{real_cond2}
\frac{\dot{a}}{a}\dot{\Phi}\gg m^2\Phi \rightarrow \frac{1}{3}\gg 1.
\end{eqnarray}
Thus, the first of these conditions, (\ref{real_cond1}), rudely contradicts the initial condition (\ref{mt<<1}), and the second condition
(\ref{real_cond2} contradicts elementary logic. Let us also notice that near cosmological singularity $t\to-\infty$ this term exponentially fast tends to infinity and at greater times $t\gg 3H_0/m^2$ it exponentially fast tends to zero $\Phi\to0$. Therefore at condition (\ref{approx_t}) and at the same time $H_0t\gg1$ we get:
\begin{equation}\label{approx_F}
\Phi\approx \Phi_0 {\displaystyle\mathrm{e}^{-\frac{m^2}{3H_0}t}}\approx \Phi_0\gg1.
\end{equation}
Let us notice that in consequence of (\ref{mt<<1}) condition (\ref{approx_t}) is much weaker than condition (\ref{approx_SM2}),
therefore the inflation solution is true for times much greater than Compton ones for scalar bosons:
\begin{equation}\label{tc}
t\ll t_{inf}=\frac{\sqrt{3}|\Phi_0|}{m}\gg \frac{1}{m}.
\end{equation}
Further, let us notice that the first term in (\ref{Phi1}) near cosmological singularity tends to infinity exponentially fast and that fact itself points at instability of the solution (\ref{Sol0}).

Let us notice that since we deal with the cosmological model being described by the non-linear system of two ordinary differential equations (\ref{d2F(t)}) and
(\ref{Einst_Eq_S}), then the approximation of slow rolling (\ref{mt<<1}) should spread to the entire system of these equations and not to single equation of the scalar field (\ref{d2F(t)}). However in this case we would have neglected the massive term in the Einstein equation i.e. instead of  (\ref{Einst_Eq_S})
we would get the following equation:

\begin{equation}\label{Einst_Eq_S_2}
3\frac{\dot{a}^2}{a^2}=8\pi\dot{\Phi}^2.
\end{equation}
Then, however, the substitution of the solution (\ref{Sol0}) in this equation would give $\dot\Phi_0=0\rightarrow a=\mathrm{Const}$, i.w. instead of inflation we would obtain a flat Minkovsky world. Thus, the solution (\ref{Sol0}) $\Phi=\Phi_0$ is mathematically incorrect even in approximation of slow rolling.

Which solution is correct in approximation of slow rolling? It is easy to find the first integral of the differential equation (\ref{d2F(t)slow}):
\begin{equation}
\label{Sol1}
a^3\dot\Phi=C_1=\mathrm{Const}.
\end{equation}
The solution (\ref{Sol0}) corresponds to particular choice of the constant $C_1= 0 $ in (\ref{Sol1}). As so often is the case in the theory of systems of non-linear ordinary differential equations, particular solutions obtained by casting out arbitrary constants turn to be instable. If we assume (\ref{Sol1}) $C_1\not=0$ and substitute the obtained result in the Einstein equation (\ref{Einst_Eq_S}), then in the approximation of slow rolling i.e. for equation (\ref{Einst_Eq_S_2}), immediately obtain the solution
\begin{equation}\label{a(t),Cnot0}
a(t)=a_1 t^{1/3};\quad a_1=C_1^{1/3}(24\pi)^{1/6},
\end{equation}
corresponding to \emph{extremely stiff} equation of state $p=\varepsilon$, i.e. $\kappa=+1$. The substitution of the found scale factor (\ref{a(t),Cnot0}) into the first integral (\ref{Sol1}) gives the final solution of the field equation in the approximation of slow rolling \cite{Ignatev_Instability}:
\begin{equation}\label{Phi_Sol1}
\Phi=C_2 +\frac{1}{\sqrt{24\pi}}\ln  t,
\end{equation}
where $C_2$  is an arbitrary constant.
\begin{thm}\label{noncorrect_solution}
Thus, the standard model is founded on the incorrect solution of the Einstein - Klein - Gordon equations in the approximation of slow rolling.
The correct solution of these equations in the approximation of slow rolling leads to finite beginning of the Universe with extremely stiff equation of state near cosmological singularity.
\end{thm}

\subsection{The Instability of the Standard Solution in the Approximation of Slow Rolling}
Let us consider perturbations of the standard model depending on time and caused by factor of mass $m^2$, which we will assume a value of the same order of smallness with perturbations $\delta(t)$ and $\phi(t)$:
\begin{eqnarray}\label{pert_a}
a(t)=a_0(t)(1+\delta(t));\\
\label{pert_f0}
\Phi(t)=\Phi_0(1+\phi(t)),
\end{eqnarray}
where $a_0(t) =a_1 \mathrm{e}^{H_0t}$, $H_0=\frac{8\pi}{\sqrt{3}}m|\Phi_0|$ is a standard solution of (\ref{inflat}).
Substituting (\ref{pert_f0}) in the scalar field equation (\ref{d2F(t)}) and expanding the obtained equation in series by smallness of $\phi,\delta,m^2$, in the linear approximation we obtain the following equation:
\begin{equation}\label{dF}
\ddot{\phi}+3H_0\dot{\phi}+m^2 =0,
\end{equation}
resolving which we find (for simplification of notation we assume $\Phi_0>0$):
\begin{eqnarray}
\phi=C_1+C_2\mathrm{e}^{-3H_0t}-\frac{m^2}{3H_0}t,\nonumber
\end{eqnarray}
herewith we should put $C_1=0$, $C_2=0$, since we are looking for not all solutions of the wave equation but only those that disappear at $m=0$.
Thus:
\begin{eqnarray}\label{df_sol}
\phi=-\frac{m^2}{3H_0}t.
\end{eqnarray}
The Einstein equation's perturbation in linear over $\delta$ approach lead to the equation:
\begin{equation}\label{pert_einst0}
\dot\delta=\frac{4\pi}{3H_0}\delta\varepsilon.
\end{equation}
From (\ref{eps}) it is seen that energy density's perturbation is quadratic over perturbations $\phi$ and $m^2$.
Substituting the solution (\ref{df_sol}) into equation (\ref{pert_einst0}), let us find with an account of (\ref{inflat}) and (\ref{F>>}) the equation on the scale factor's perturbation:
\begin{equation}\label{eqdelta}
\dot{\delta}=\frac{4\pi\Phi^2_0}{3H_0}\bigl(\frac{m^4}{9H^2_0}+m^2\bigr)\equiv \frac{H_0}{2}\bigl(1+\frac{1}{72\pi\Phi^2_9}\bigr)\approx \frac{H_0}{2}
\end{equation}
and hence find:
\begin{equation}
\delta=\frac{1}{2}H_0t.
\end{equation}
In order inflation to develop it is necessary that $H_0t\gg1$, thus during the inflation period \emph{relative} perturbation of the scale factor becomes a big value hence standard solution is instable. This means instability of the inflation solution with respect to factor of mass.
\begin{thm}
The inflation solution of the Einstein equations (\ref{p=-e}) -- (\ref{inflat}) is instable with respect to factor of mass:
perturbations of metrics and scalar field grow till infinity in the infinite past.
\end{thm}
%
\section{The Qualitative Analysis of the Dynamic System of the Standard Cosmological Scenario}
All the above-listed contradictions being resulting consequences of application of the approximation of slow rolling to the cosmological system of Einstein - Klein - Gordon equations let us think about necessity of more detailed mathematical analysis of this system of equations without applying the approximation of slow rolling. Basically, this is not a so complex autonomous dynamic system from the standpoint of the theory of differential equations. Therefore it makes sense to carry out its analysis using methods of the qualitative theory of ordinary differential equations. In order to do that it is necessary to make all variables included in these equations dimensionless and reduce the obtained equations to normal form i.e. to the system of differential equations of the first order, resolved relative to the derivatives.
\subsection{Reducing of the System of Equations to the Canonical Form}
First, let us carry out scaling of the equations (\ref{d2F(t)}) and (\ref{Einst_Eq_S}), transiting to dimensionless time variable $\tau$
\begin{equation}\label{tau}
\tau=mt,\Rightarrow \dot{f}=mf',
\end{equation}
where $f'=df/d\tau$. Thus, instead of equations (\ref{d2F(t)}) and (\ref{Einst_Eq_S}) we obtain:
\begin{eqnarray}
\label{Phi''}
\Phi''+3h\Phi'+\Phi=0;\\
\label{Einst_h}
3h^2=8\pi \bigl(\Phi'^2+\Phi^2\bigr),
\end{eqnarray}
where $h(\tau)$ is a Hubble constant measured in units of the Compton time:
\begin{equation}\label{h}
h(\tau)=\frac{a'}{a}=\frac{H}{m}.
\end{equation}
Let us notice that equations (\ref{Phi''}) and (\ref{Einst_h}) represent a system of ordinary non-linear differential equations which taking into account
\begin{equation}\label{dota>0}
\dot{a}\geq 0 \Leftrightarrow H\geq 0
\end{equation}
and using a standard substitution can be reduced to the form of \emph{normal autonomous system of ordinary differential equations on a plane}:
\begin{eqnarray}\label{Z}
\Phi'&=&Z;\\
\label{Z'}
Z'&=&-3hZ-\Phi,
\end{eqnarray}
where function $h(\Phi,Z)$  is \emph{algebraically} defined from the Einstein equation with a help of function $\Phi(\tau)$ and $Z(\tau)$:
\begin{equation}\label{h(tau)}
h=\sqrt{\frac{8\pi}{3}}\sqrt{Z^2+\Phi^2}.
\end{equation}
Thus, we finally obtain the system of autonomous differential equations on the plane $(\Phi,Z)$:
\begin{equation}\label{eqs}
\left\{\begin{array}{l}
\Phi'=Z;\\[12pt]
Z'=-\sqrt{24\pi}\sqrt{Z^2+\Phi^2}Z-\Phi.
\\
\end{array}\right.
\end{equation}
Let us rewrite this system in terms of the standard theory of dynamic systems (see e.g. \cite{Bogoyavlensky}):
\begin{equation}
\left\{\begin{array}{l}
\displaystyle\frac{dx}{dt}=P(x,y); \\[12pt]
\displaystyle\frac{dy}{dt}=Q(x,y),\\[12pt]
\end{array}\right.
\end{equation}
where $x=\Phi$; $y=Z$;
\begin{equation}\label{PQ}
P(x,y)=y;\; Q(x,y)=-\sqrt{24\pi}\sqrt{x^2+y^2}Z-x.
\end{equation}
This system of equations can be investigated using the qualitative theory of differential equations and asymptotic behavior of the solutions at  $t\to\pm\infty$ can be defined. The next property of the SCM is quite important one.
\begin{thm}\label{utv6}
The evolution of the Universe in the SCM in terms of time variable $\tau$ does not depend on any parameters and is fully defined by initial conditions.
\end{thm}
Therefore the results of the qualitative analysis of the SCM equations have general character.

\subsection{Singular Points of the Dynamic System}
Singular points of the dynamic system $M_0(x_0,y_0$ (\ref{eqs}) are defined by zeroes of the derivatives (see e.g. \cite{Bogoyavlensky}):
\[P(x_0,y_0)=0;\quad Q(x_0,y_0)=0.\]
It is not difficult to see that the dynamic system (\ref{eqs}) has a unique singular point:
\begin{equation}\label{Phi0,Z0}
M_0=(0,0)\longleftrightarrow \Phi_0=0;\; Z_0=0.
\end{equation}
\subsection{A Kind of Singular Point}
To define a kind of a singular point it is necessary to find the eigenvalues of the characteristic polynomial:
\begin{equation}\label{haracter_pol}
\Delta(\lambda)=\left|
\begin{array}{ll}
P'_x(x_0,y_0)-\lambda & P'_y(x_0,y_0)\\[12pt]
Q'_x(x_0,y_0) & Q'_y(x_0,y_0)-\lambda\\
\end{array}
\right|=0,
\end{equation}
where partial derivatives of functions $P(x,y),Q(x,y)$ are calculated in a singular point $M_0$. Calculating derivatives of functions $P,Q$ (\ref{PQ}), let us find:
\begin{eqnarray}
P'_x(0,0)=0;& P'_y(0,0)=1;\nonumber\\
Q'_x(0,0)=-1;& Q'_y(0,0)=0.\nonumber
\end{eqnarray}
Thus, the characteristically polynomial (\ref{haracter_pol}) is equal to:
\[
\Delta(\lambda)=\left|
\begin{array}{rr}
-\lambda & 1\\[12pt]
-1 & -\lambda\\
\end{array}
\right|=0,
\]
where from we find its roots
\begin{equation}\label{lambda}
\lambda=\pm i.
\end{equation}
Since the eigenvalues turned to be pure imaginary ones then \emph{the single singular point} (\ref{Phi0,Z0}) of the dynamic system (\ref{eqs}) is its center
(see \cite{Bogoyavlensky}). In such a case at $\tau\to+\infty$ a phase trajectory of the dynamic system is winded round this center making an infinite number of turns.
\begin{thm}
Phase space f the dynamic system founded on the equation of classical scalar field (\ref{d2F(t)}) and the Einstein equation (\ref{Einst_Eq_S}), has a single center (\ref{Phi0,Z0}), where:
\begin{equation}\label{center}
t\to+\infty \Rightarrow \Phi\to0;\quad \dot{\Phi}\to 0\Rightarrow H\to 0.
\end{equation}
Thus, in spite of widespread notion the system does not come to the mode of constant inflation but vice versa its expansion stops and the Universe becomes flat (see Fig. \ref{ris1}).
\end{thm}
\Fig{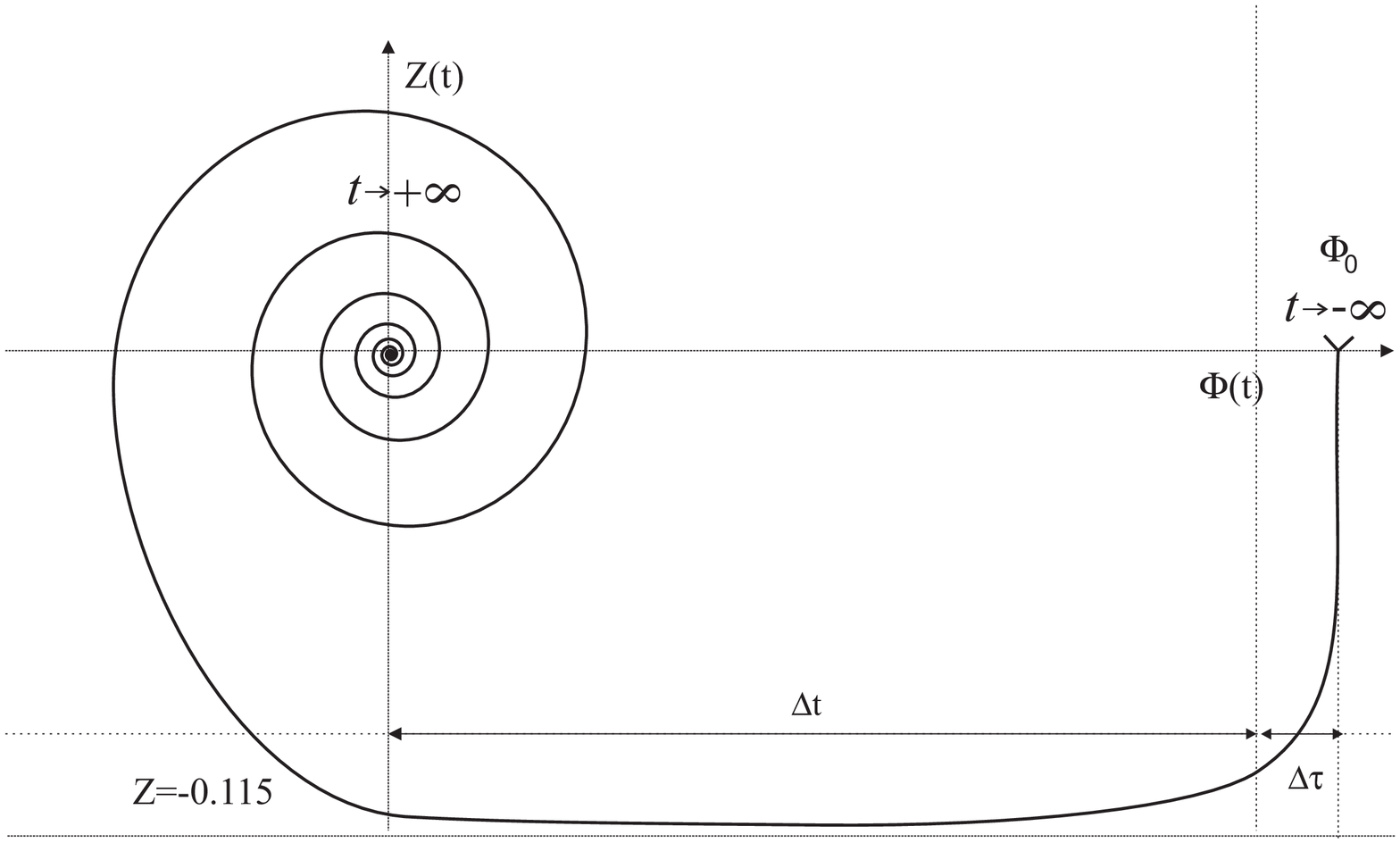}{12}{\label{ris1}The qualitative view of the phase trajectory of the dynamic system (\ref{eqs}). On this figure $\Delta\tau$ is a characteristic time of decrease of the potential's rate of change till the ``bottom'' of the plot, $Z_0\approx-0.115$, $\Delta t$ is a characteristic time of decrease of the potential's value with a constant speed $\Phi'\approx Z_0$. After this instant, the winding of the phase space round the zero center begins. The number of turns of the spiral at that is infinite.}
\section{Phase Planes of the Dynamic System (\ref{eqs})}
 It needs to keep in mind that time variable on all plots is $\tau$, i.e. time measured on Compton scale. Since in consequence of statement \ref{utv6} the evolution of the investigated dynamic system (\ref{eqs}) is defined only by initial conditions let us investigate the dependency of the evolution details on initial conditions. Further, let us take an assumption that conforms with SCM $\dot{\Phi}_0=0$.  Let us first show the results of numerical simulation of phase planes of the dynamic system (\ref{eqs}) at great initial values of the potential $\Phi_0\geq 10$.  Let us recall that one of necessary conditions of validity of the approximation of slow rolling is $\Phi_0\gg1$ (\ref{approx_F}). Let us investigate the properties of the phase trajectory's in terms of plot on Fig. \ref{ris1}:
 \begin{enumerate}
 \item The initial stage with duration $\Delta\tau$ ñ $\Phi\approx\Phi_0$ is on the right side of the plot; this stage is characterized by fast decrease of the derivative from $0$ till  ``enigmatic number'' $-0.115$. Actually there is no any enigma in this number (see equation (\ref{eqs})):
\begin{equation}\label{Z0}
Z_0=-\frac{1}{\sqrt{24\pi}}\approx - 0.1151647165.
\end{equation}
 Actually, inflation happens at this stage.
 \item The Middle stage with duration $\Delta t$ is a medium part of the plot; $Z=\Phi'\approx \mathrm{Const}=Z_0$ at this stage. Potential falls to significantly small values at this stage.
\item The Final stage of the evolution with infinite duration; Decaying oscillations of the potential and its derivative happen at this stage.
The University becomes asymptotically flat.
\end{enumerate}
On the plots below there are shown the results of numerical simulation of the dynamic system (\ref{eqs}) at various initial conditions. In this paper it is used a Rosenbrock's method well adapted to integration of stiff systems of differential equations.
\subsection{Initial conditions: $\Phi(-1000)=10,\\ \dot{\Phi}(-1000)=0$}
\Fig{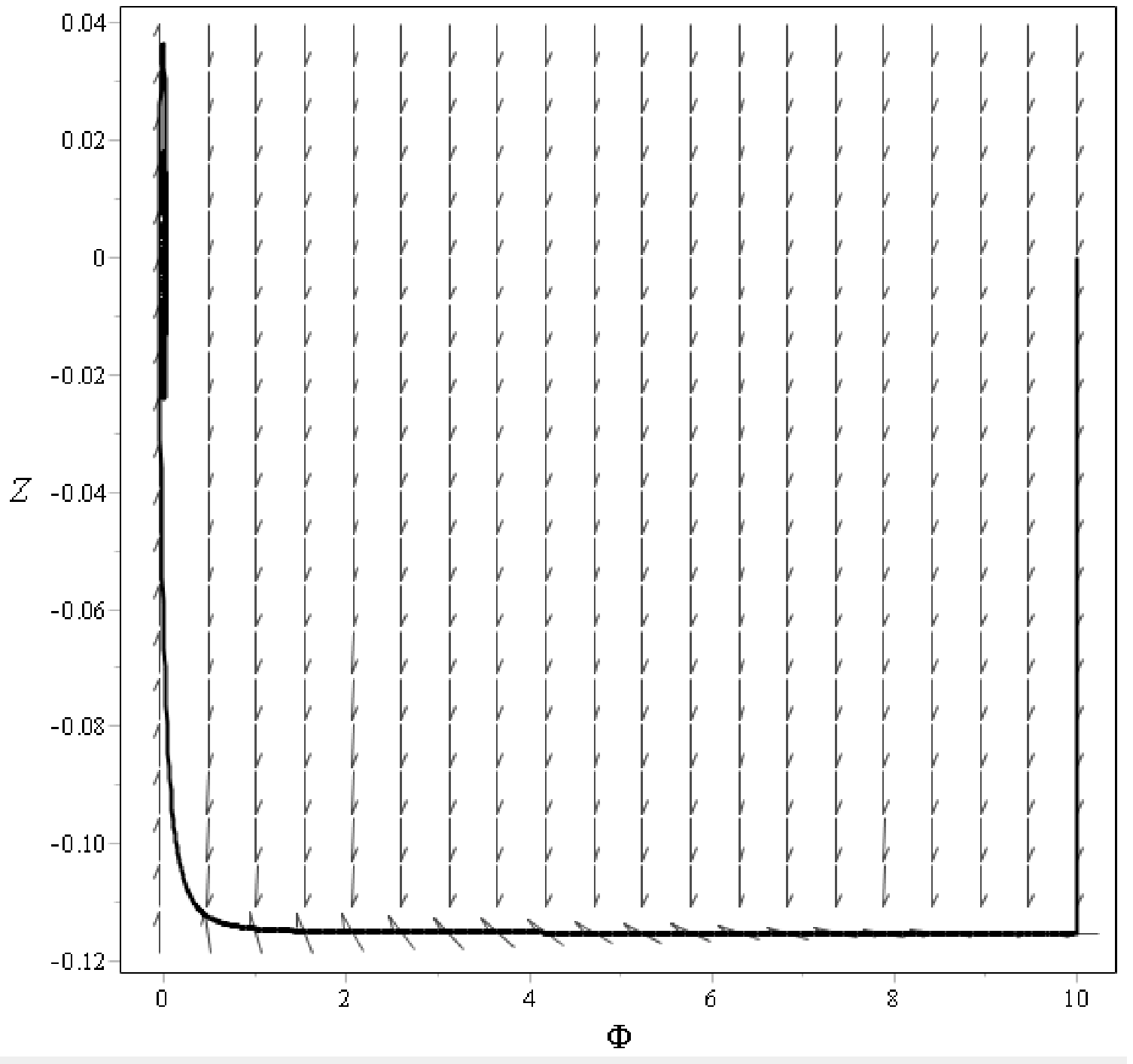}{7}{\label{ris2}The large-scale phase plane of the dynamic system (\ref{eqs}) $\tau\in[-1000,1000]$.}

\Fig{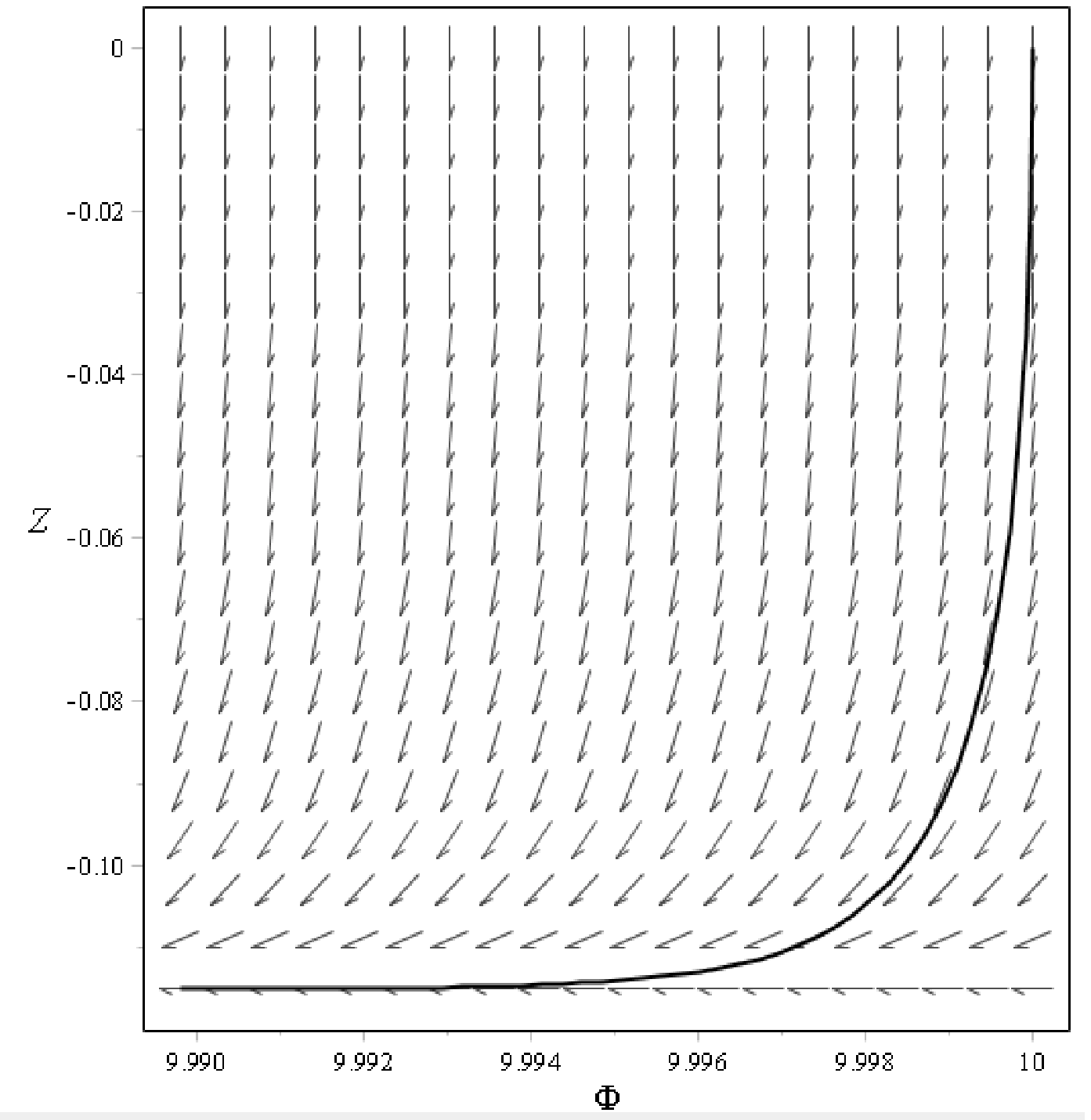}{7}{\label{ris3}The initial stage of descent of the dynamic system (\ref{eqs}) (the right-most part of the plot on Fig. \ref{ris1}) $\tau\in[-1000,-999.9]$; $\Delta\tau\lesssim 10^{-1}$.}
\Fig{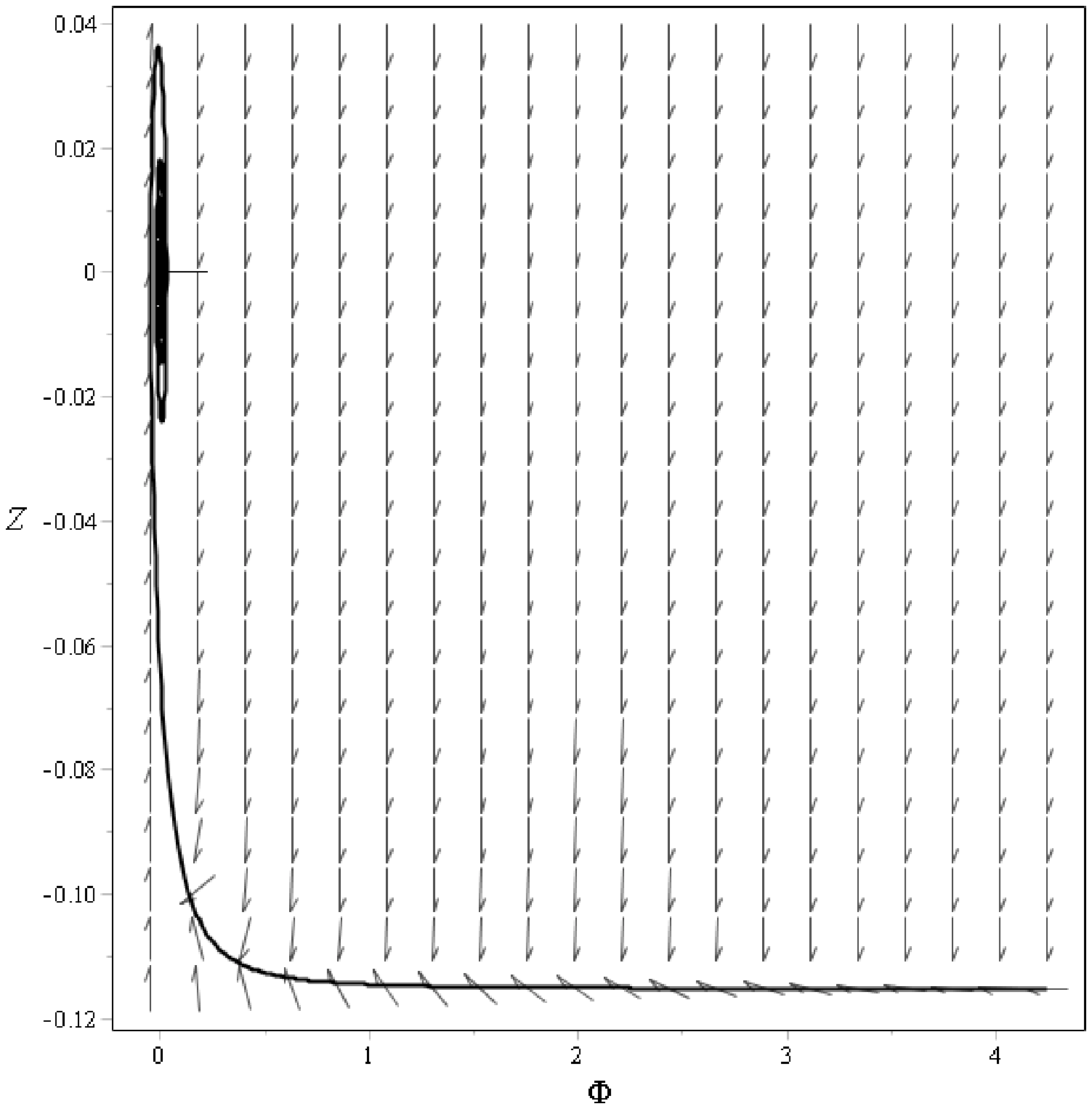}{7}{\label{ris4}The middle stage of the dynamic system (\ref{eqs}) $\Phi'\approx \mathrm{Const}\approx -0.115$ $\tau\in[-950,100]$.}
Since characteristic details of the system's phase planes that are interesting for us (\ref{eqs}) have incomparable scales, we show fragments of phase planes on different time intervals.
\Fig{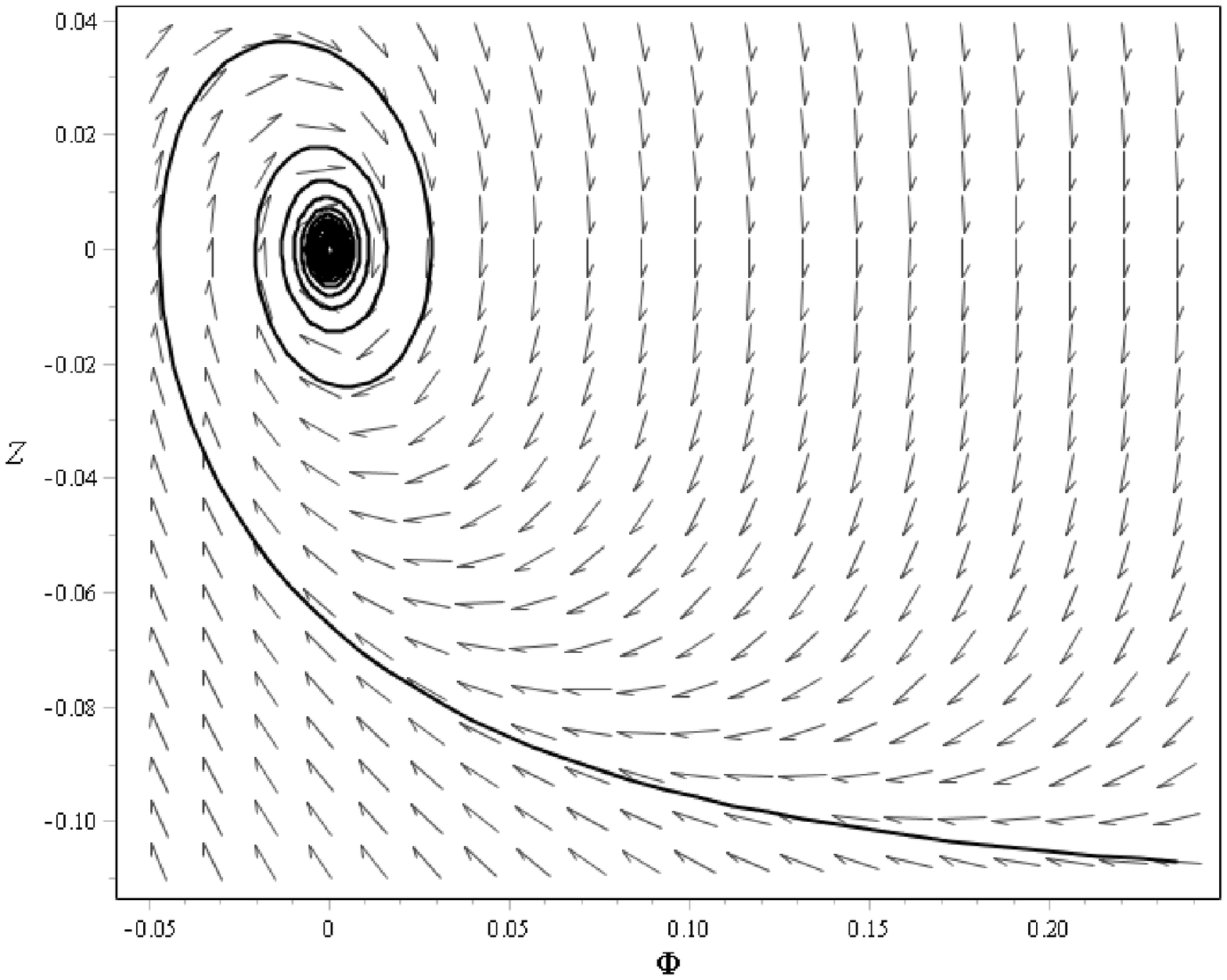}{7}{\label{ris5}Winding round the center $M_0=(0,0)$ of the dynamic system (\ref{eqs}) (the left-most part of the plot shown on Fig. \ref{ris1}) $\tau\in[-915,-700]$. }
\Fig{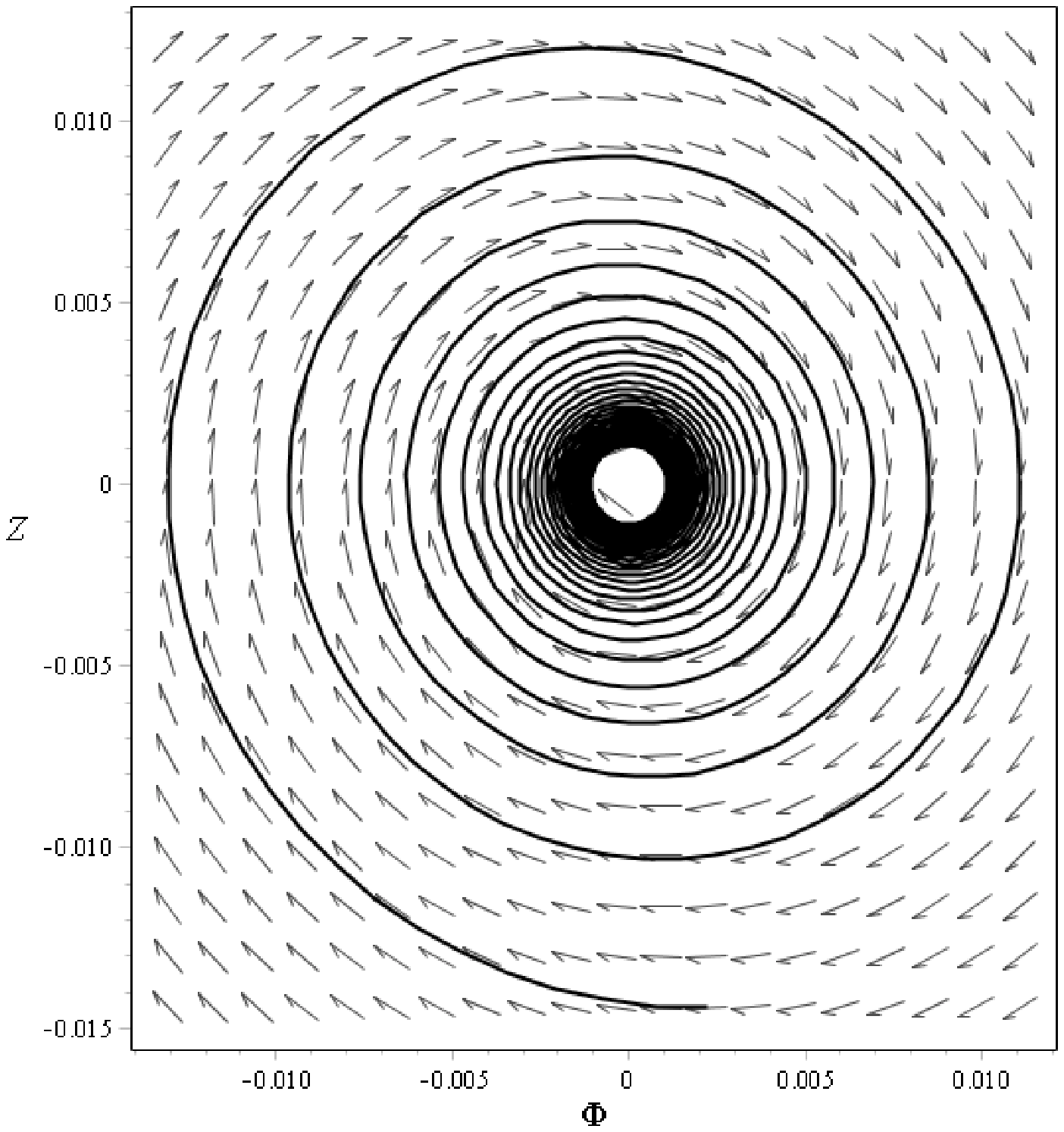}{7}{\label{ris6}The final stage of the dynamic system (\ref{eqs}): winding round the center $M_0=(0,0)$ at initial conditions: $\Phi(-1000)=10,\dot{\Phi}(-1000)=0$; $\tau\in[-900,-700]$.}
\Fig{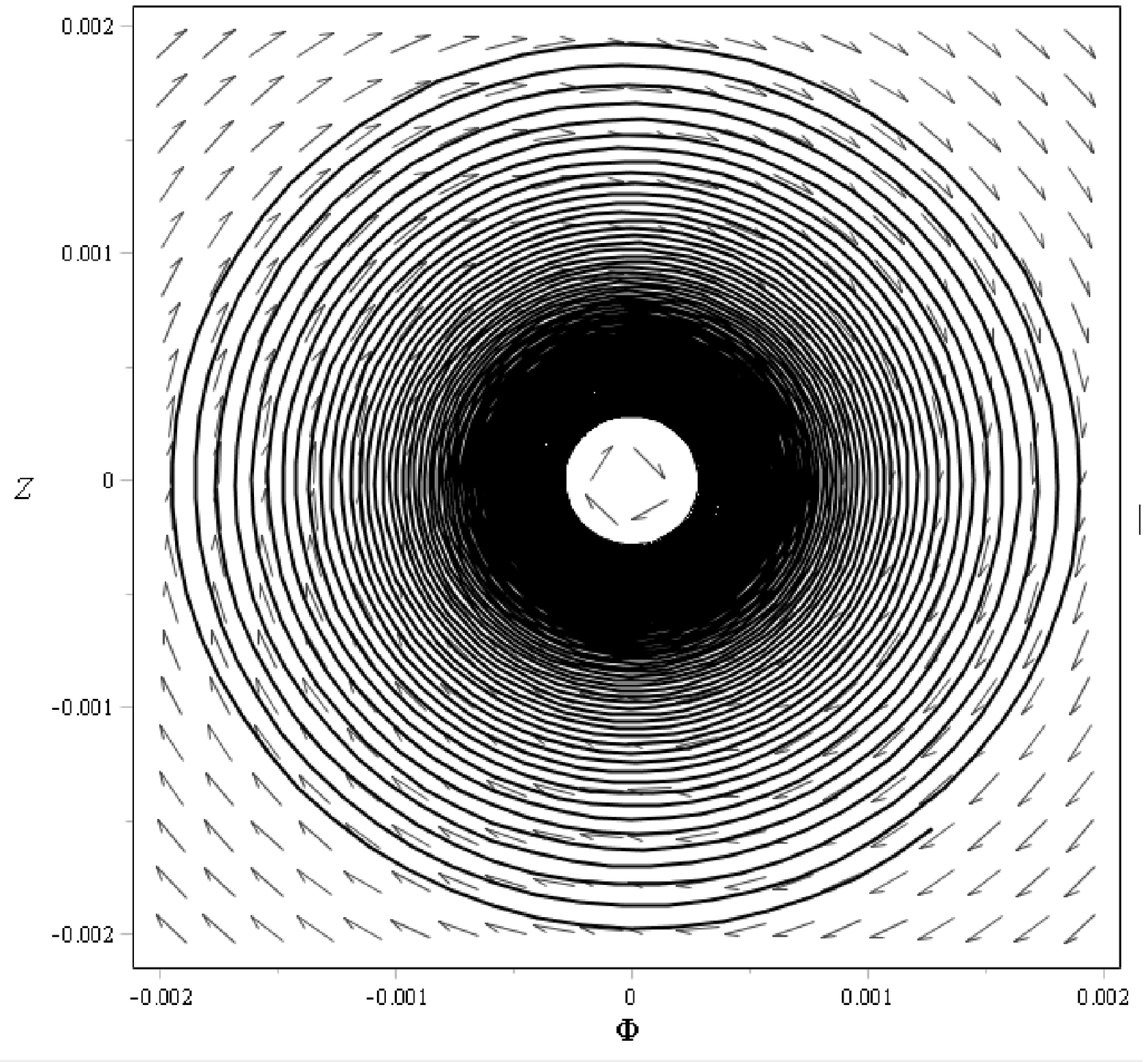}{7}{\label{ris7}The final stage of the dynamic system (\ref{eqs}): winding round the center $M_0=(0,0)$ of the dynamic system (\ref{eqs}) (left-most part of the plot shown on Fig. \ref{ris1}) $\tau\in[-800,-100]$. }
%
\subsection{Initial conditions: $\Phi(-1000)=0.1,\dot{\Phi}(-1000)=0$}
\Fig{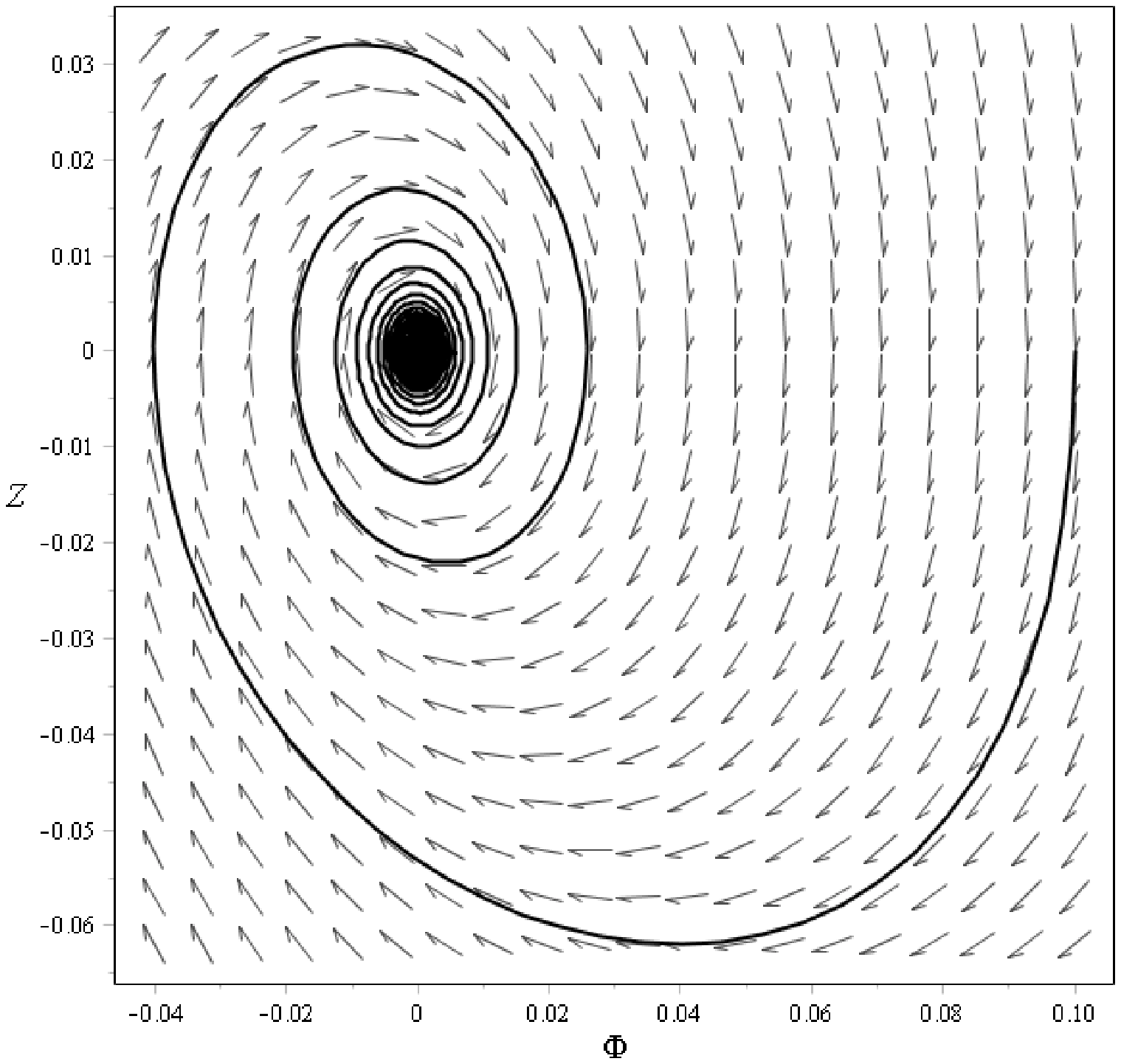}{7}{\label{ris12}Large-scale phase plane of the dynamic system (\ref{eqs}) $\tau\in[-1000,1000]$.}
\Fig{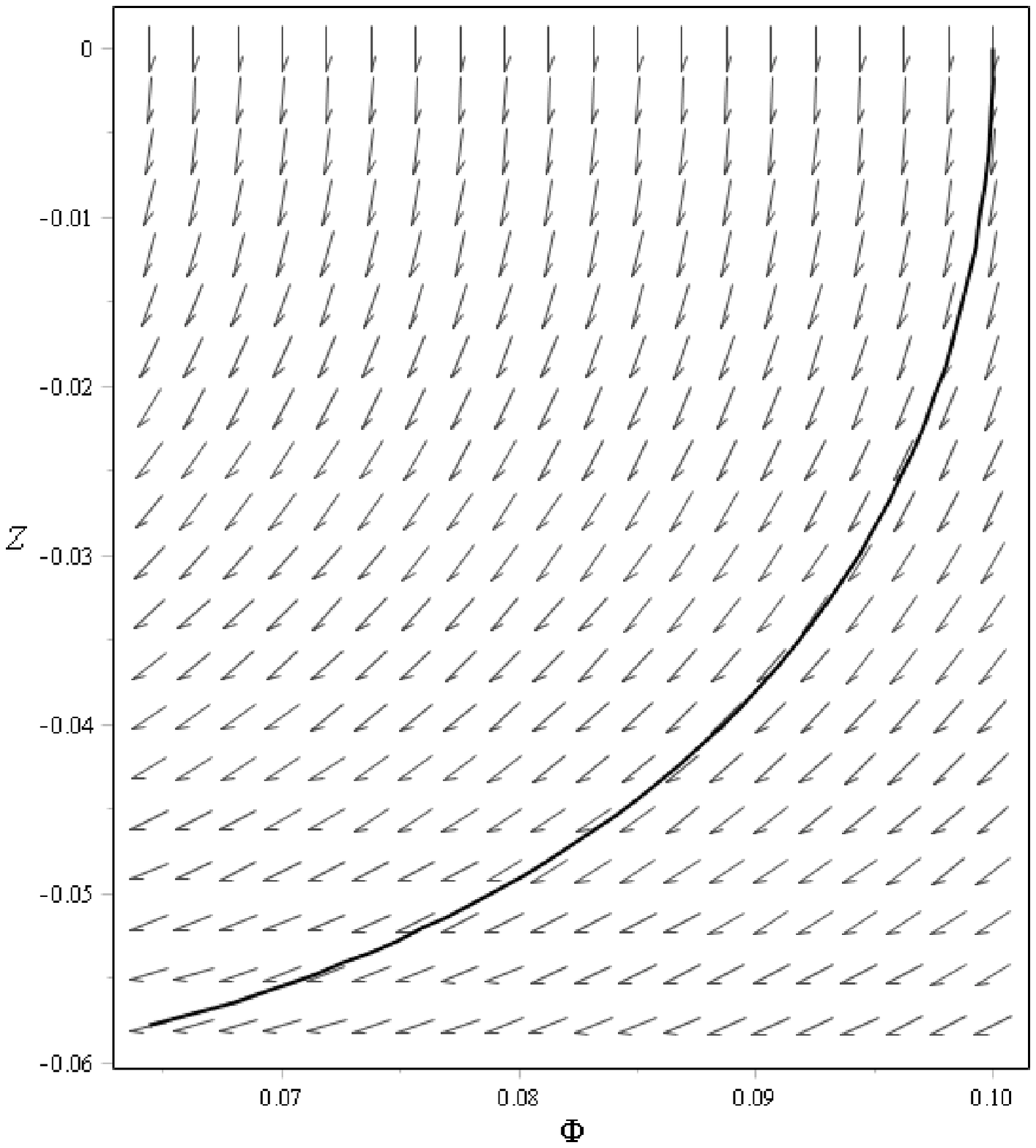}{7}{\label{ris13}The initial stage of descent of the dynamic system (\ref{eqs}) (the right-most part of the plot on Fig. \ref{ris1}) $\tau\in[-1000,-999]$; $\Delta\tau\lesssim 1$.}
\Fig{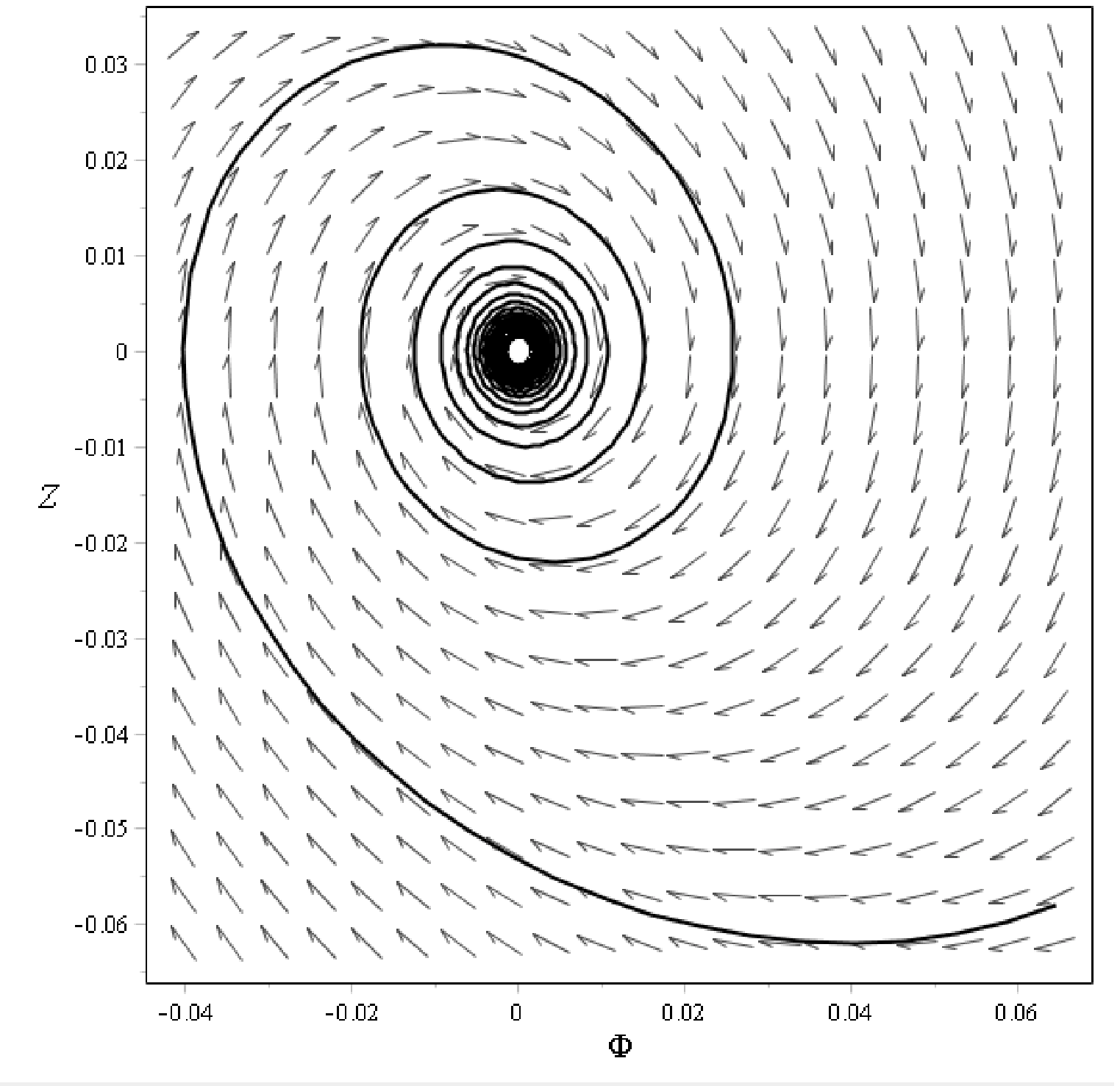}{7}{\label{ris14}The middle stage of the dynamic system (\ref{eqs}) $\Phi'\approx \mathrm{Const}\approx -0.115$ $\tau\in[-999,-850]$.}
\Fig{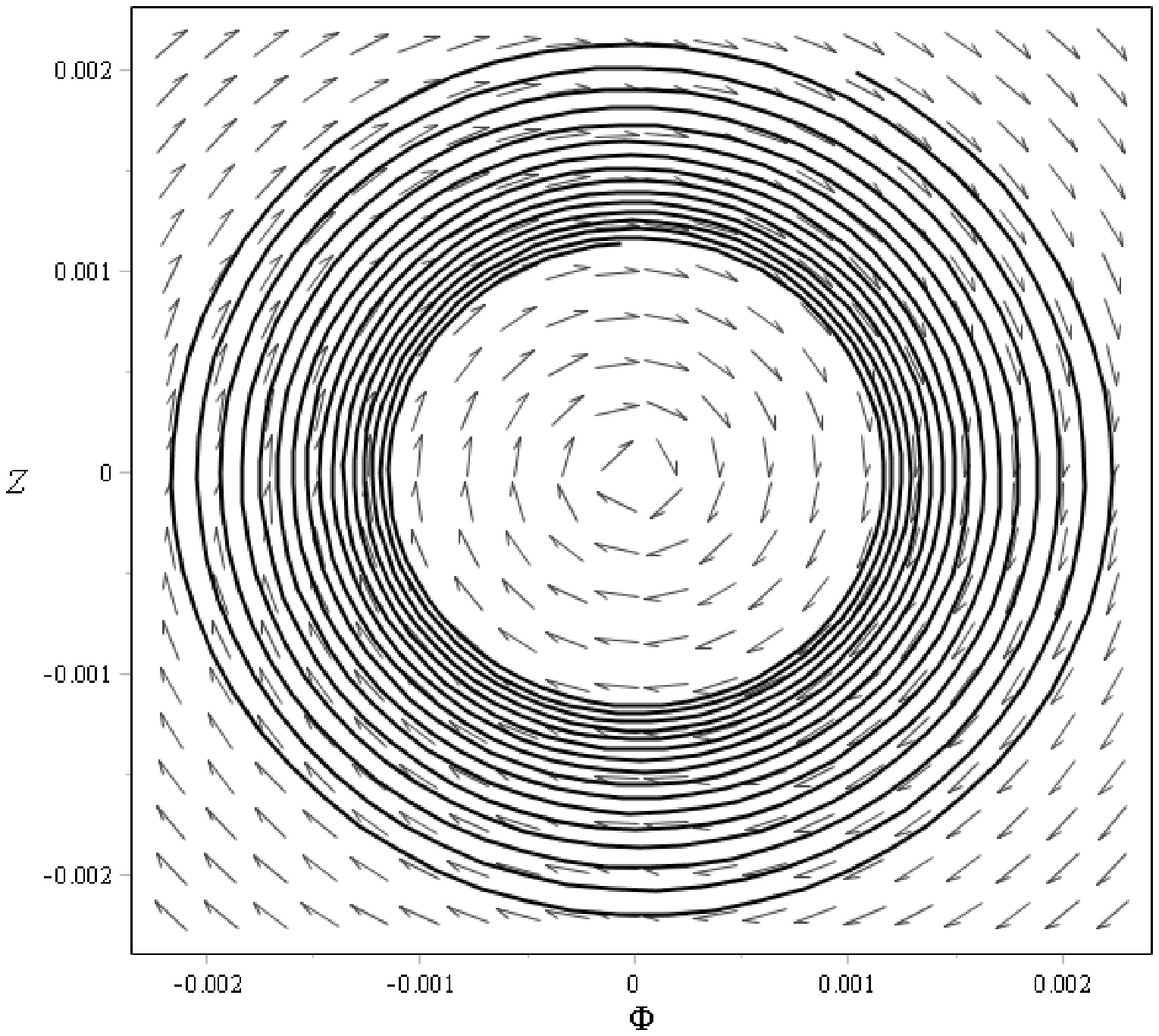}{7}{\label{ris15}Winding round the center $M_0=(0,0)$ of the dynamic system (\ref{eqs}) (left part of the plot shown on Fig. \ref{ris1}) $\tau\in[-900,-800]$. }
%
\section{Numerical Integration of the Dynamic\\ Equations}
\subsection{Evolution of the Potential and its Derivative}
Phase planes of the dynamic system (\ref{eqs}), shown on Fig. \ref{ris2} -- \ref{ris15} do not provide information about certain details of the cosmological evolution; these details only can be obtained using direct numerical integration of the original system of Einstein - Klein - Gordon  equations.
Let us see the results of numerical integration of these equations using Rosenbrock's method.
First of all let us see the large-scale phase plane of the middle and final stages of the cosmological model's evolution, which is obtained by direct numerical integration of the system of equations (\ref{eqs}) and, practically, fully reproduces a sketch of phase plane on Fig. \ref{ris1}.
\FigReg{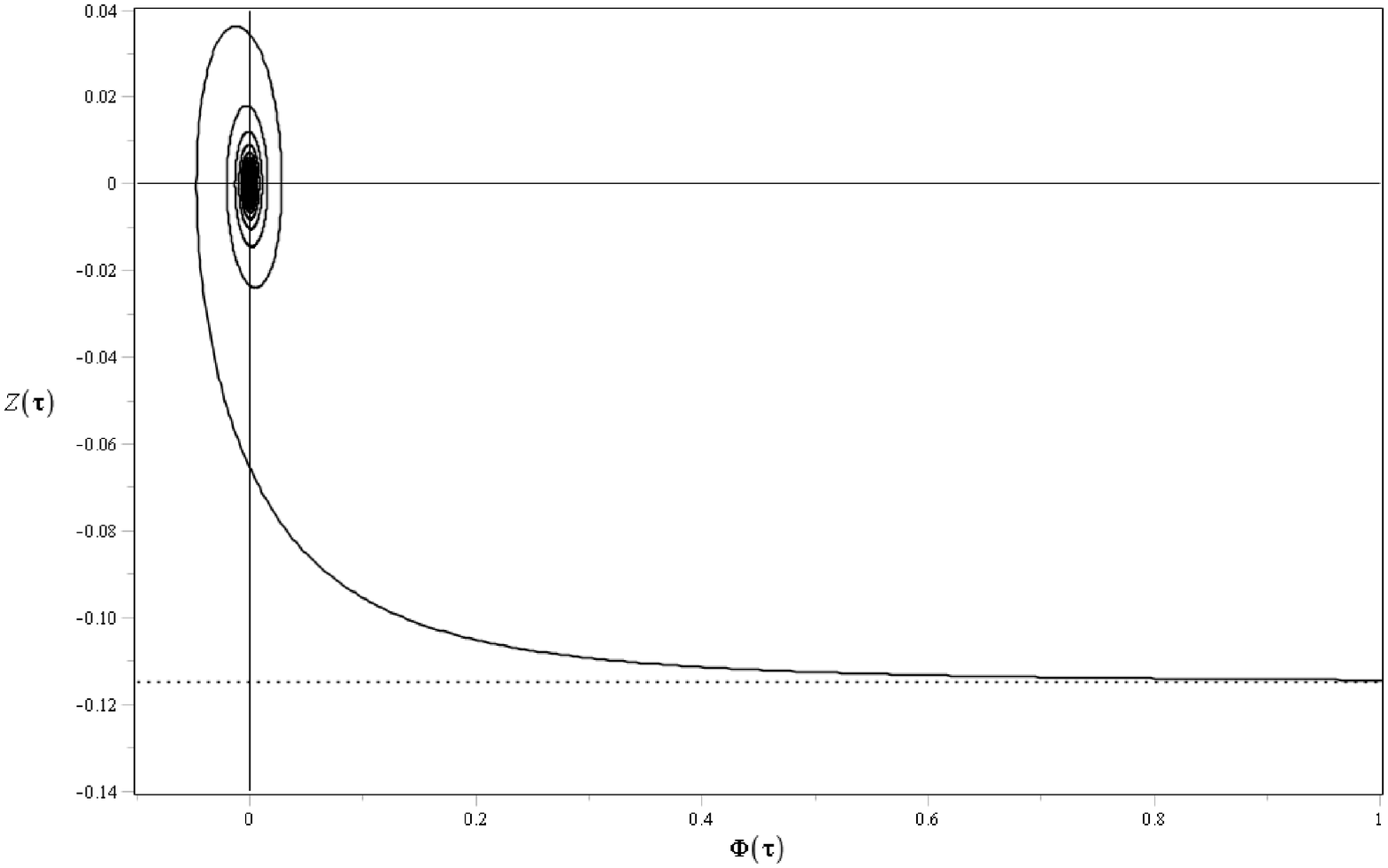}{7.7}{3.85}{\label{ris16}Phase plane of the dynamic system (\ref{eqs}), obtained by direct numerical integration using Rosenbrock's method of the system (\ref{eqs}) with initial conditions: $\Phi(-100)=100;\dot{\Phi}(-100)=0$; $\tau\in [640,1000]$.}
Then, let us see the plots of functions $\Phi(\tau)$ è $Z(\tau)$. These plots, considering their properties, should also be viewed at different scales and intervals.
\Fig{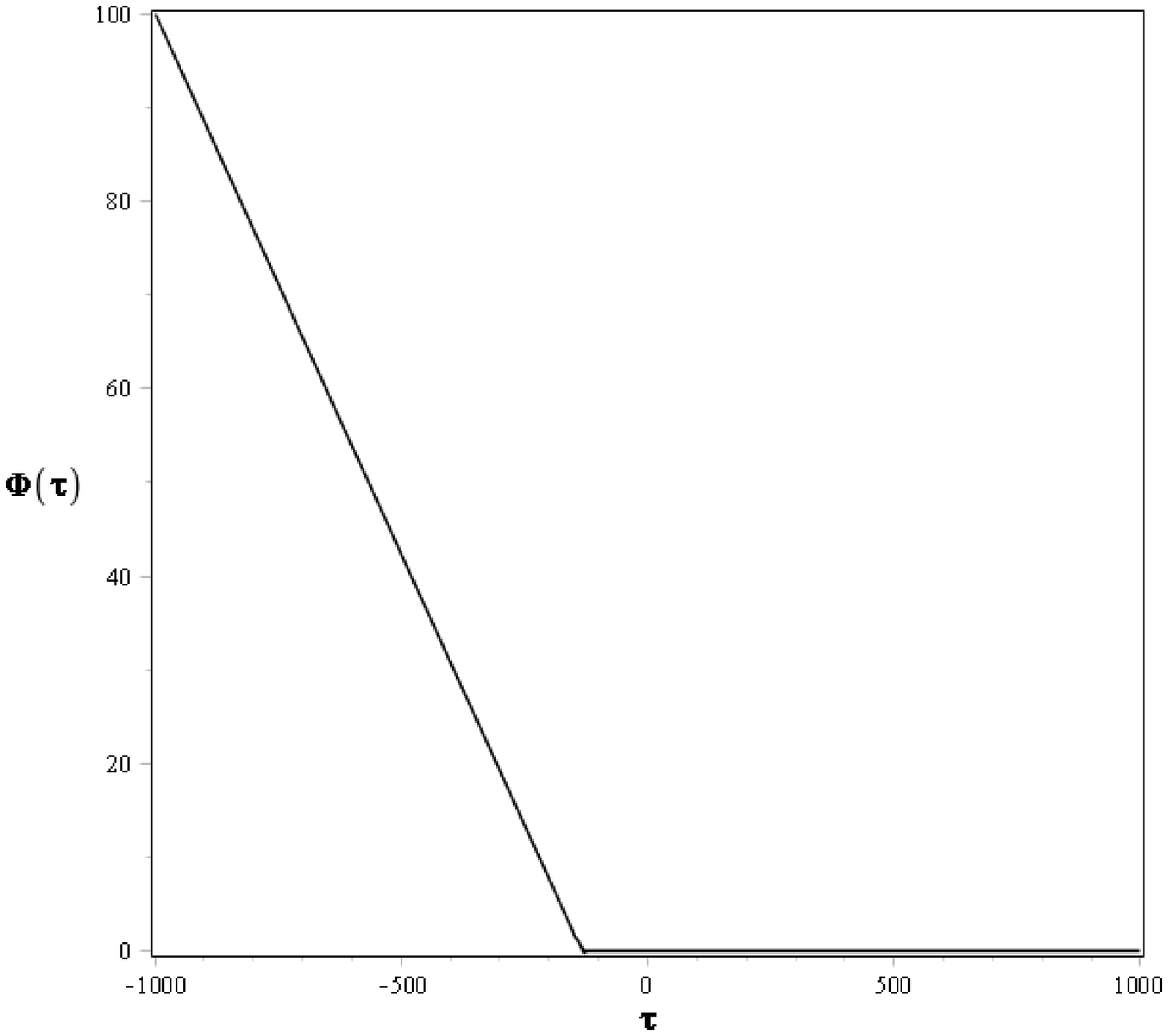}{7}{\label{ris17}Large-scale behavior of the potential $\Phi(\tau)$ at initial conditions $\Phi(-1000)=100,\dot{\Phi}(-1000)=0$;
$\tau\in[-1000,1000]$.}
As is seen from the plots below the value of the potential falls virtually linearly with time on interval \\ $\tau\in[-1000,-100]$ by approximate law:\\
\[\Phi(\tau)=\Phi_0-0.1 (\tau-\tau_0)\Delta\tau.\]
By the way, this linear fall corresponds to approximate law $\dot{\Phi}=\mathrm{Const}$ and not to $\Phi=\mathrm{Const}$ which would correspond the case of SCM. From $\tau\approx -100$ linear fall of the potential stops and further the potebntial evolves in accordance with the decaying oscillations law.
\Fig{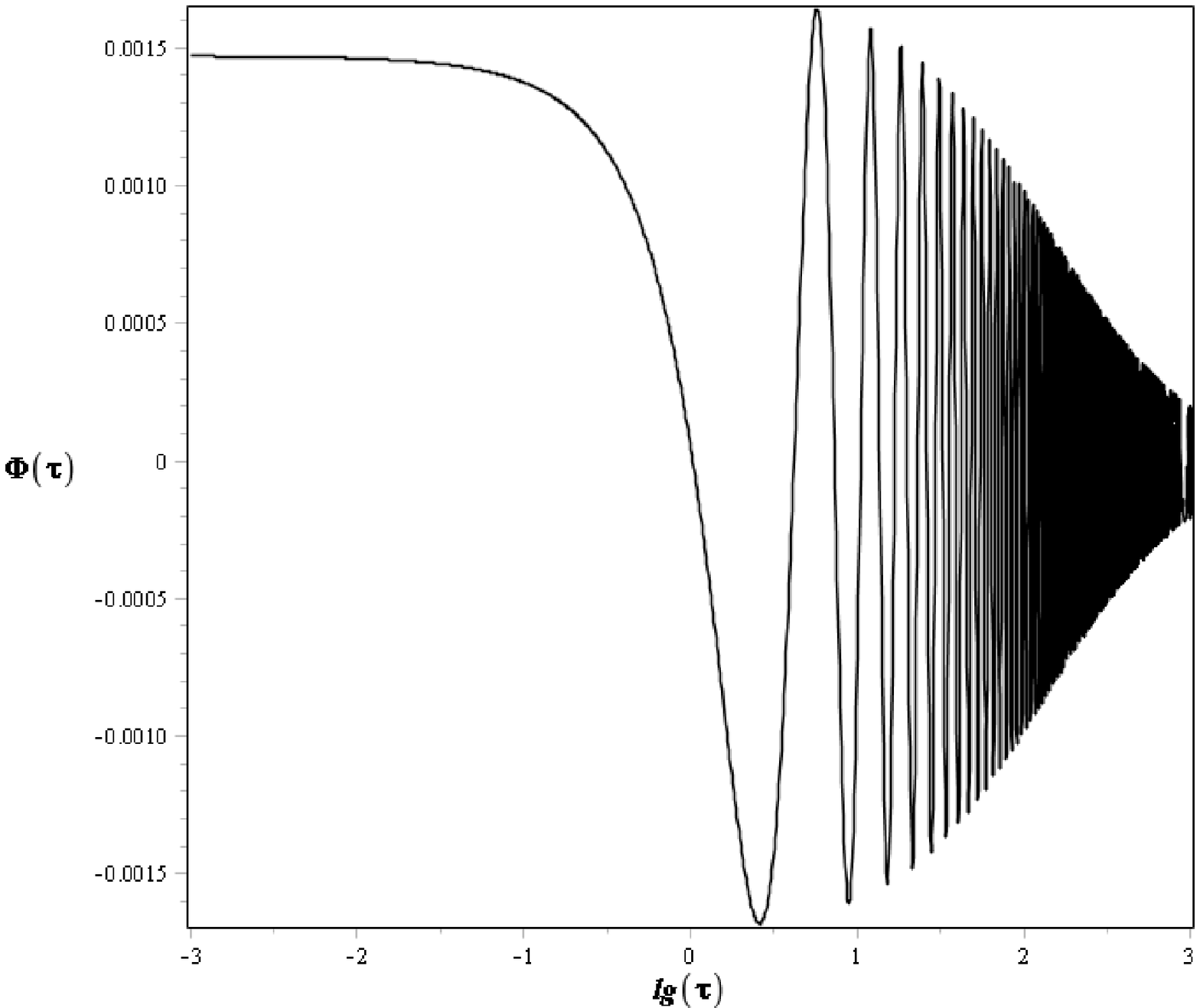}{7}{\label{ris18}The final stage of the evolution of the scalar potential $\Phi(\tau)$ at initial conditions $\Phi(-1000)=100,\dot{\Phi}(-1000)=0$ in the logarithmic time scale; $\tau\in[0.001,1000]$.}
\Fig{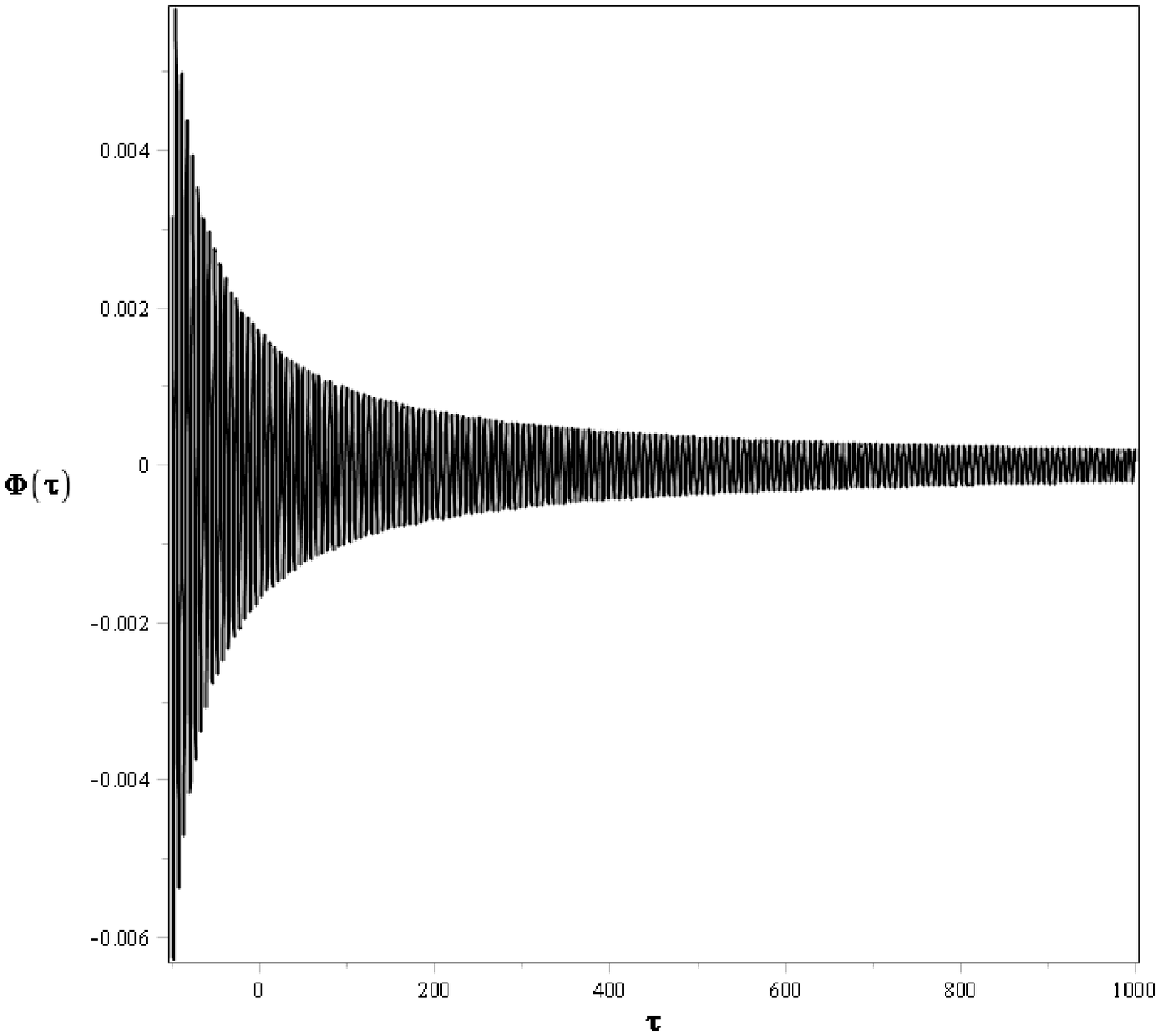}{7}{\label{ris19}Transition of the plot of the potential $\Phi(\tau)$ to the oscillating regime at initial conditions $\Phi(-1000)=100,\dot{\Phi}(-1000)=0$; $\tau\in[-100,1000]$.}
\Fig{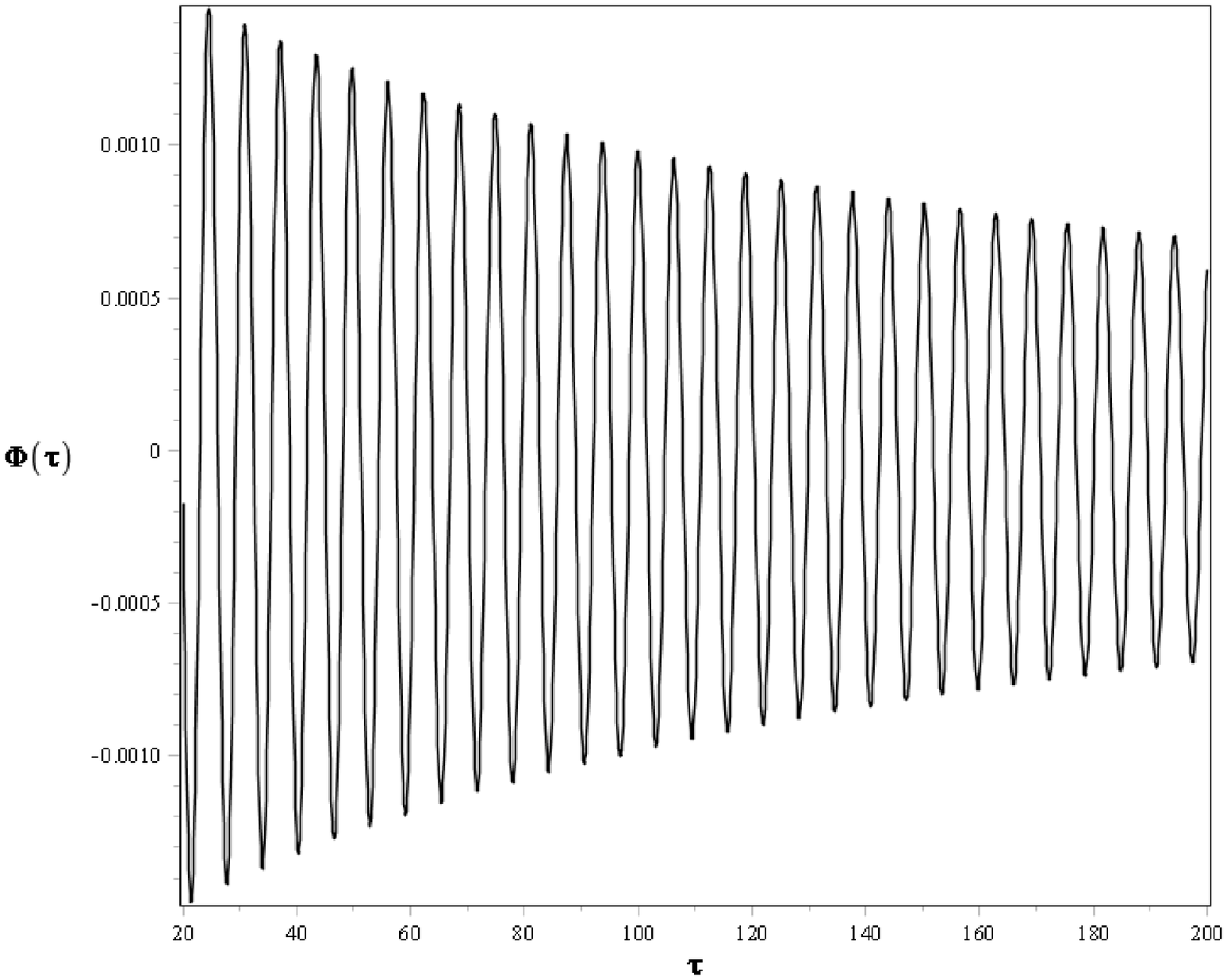}{7}{\label{ris20}Regime of decaying oscillations of the scalar potential $\Phi(\tau)$ at initial conditions $\Phi(-1000)=100,\dot{\Phi}(-1000)=0$; $\tau\in[20,200]$.}
On the following plots the identical regimes of the potential's derivative are shown.
\Fig{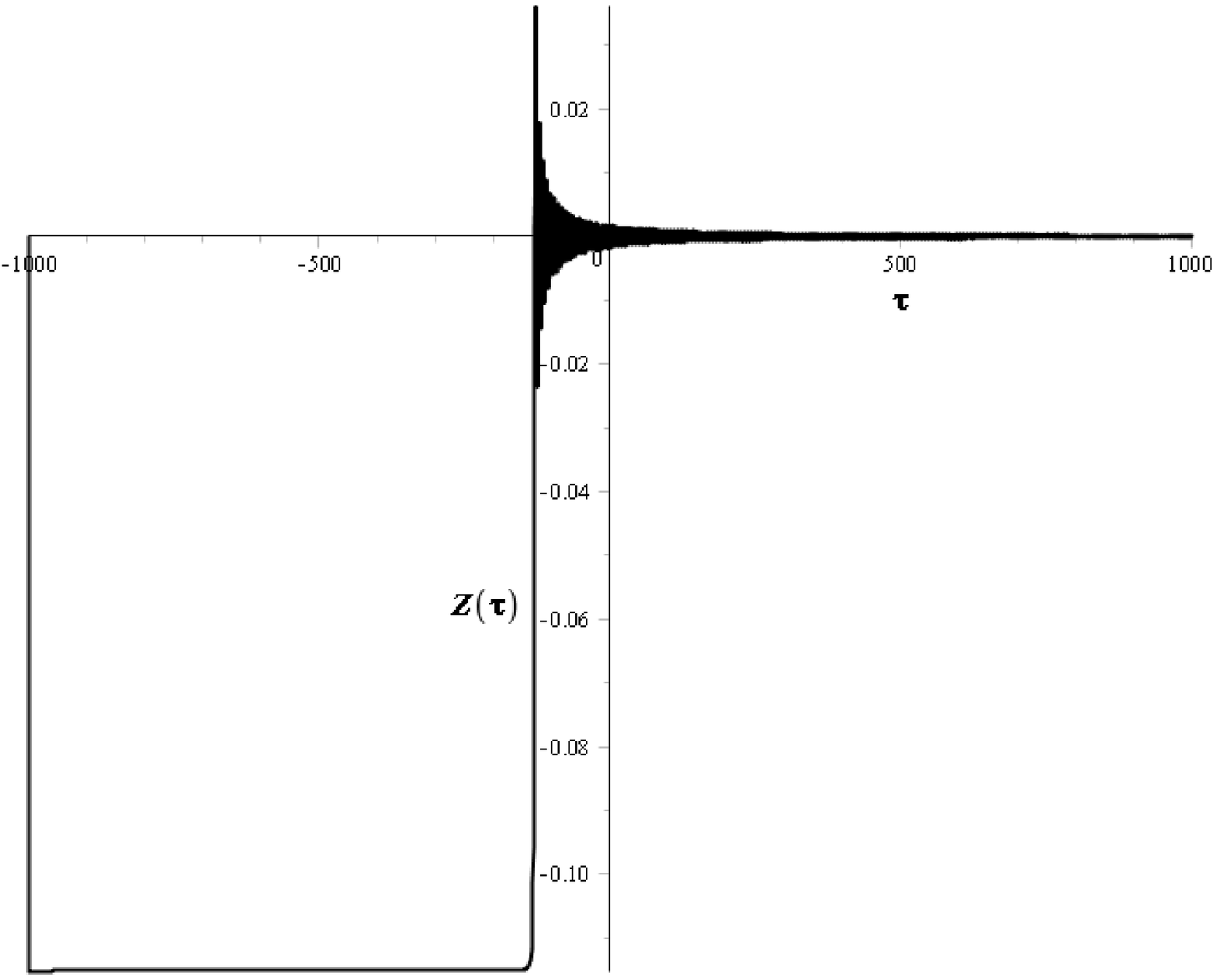}{7}{\label{ris21}Large-scale behavior of the potential's derivative $Z(\tau)=\Phi'(\tau)$ at initial conditions $\Phi(-1000)=100,\dot{\Phi}(-1000)=0$; $\tau\in[-1000,1000]$. At the bottom of the plot it is $\Phi'(\tau)\approx Z_0$ in accordance with  (\ref{Z0}).}
The regime of linear fall of the potential on Fig. \ref{ris17}. corresponds to the bottom of the plot $Z\approx -0.115$.
\Fig{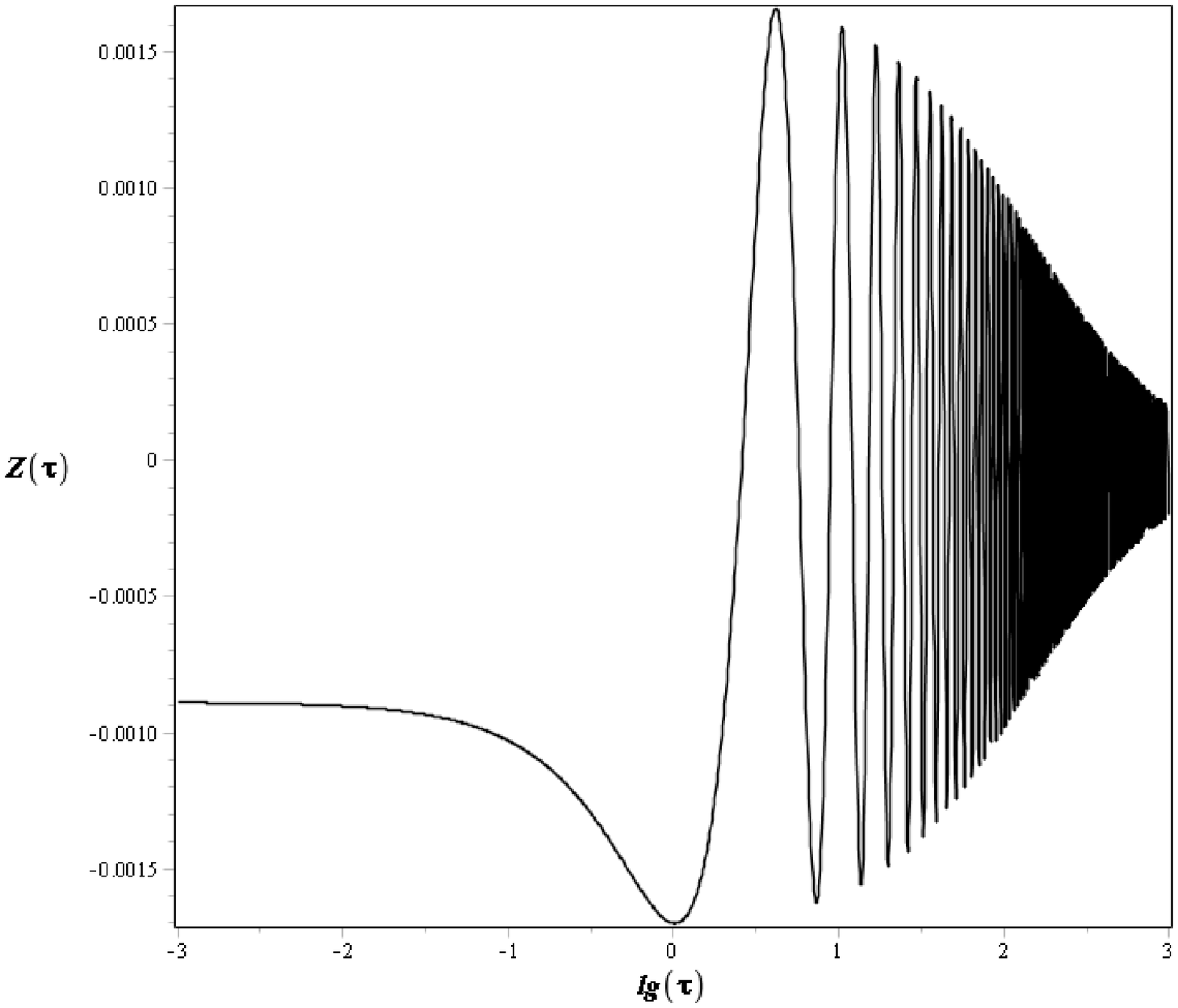}{7}{\label{ris22}The final stage of the scalar potential's evolution $Z(\tau)=\Phi'(\tau)$ at initial conditions $\Phi(-1000)=100,\dot{\Phi}(-1000)=0$ in the logarithmic time scale; $\tau\in[0.001,1000]$.}
\subsection{The Evolution of the Hubble Constant, Cosmological Acceleration and Scale Factor}
The Hubble ``constant'' $H(t)$ is connected by simple relation (\ref{h}) with a <<constant>> $h(\tau)$ normalized on mass of the scalar field and calculated through (\ref{h(tau)}).

\Fig{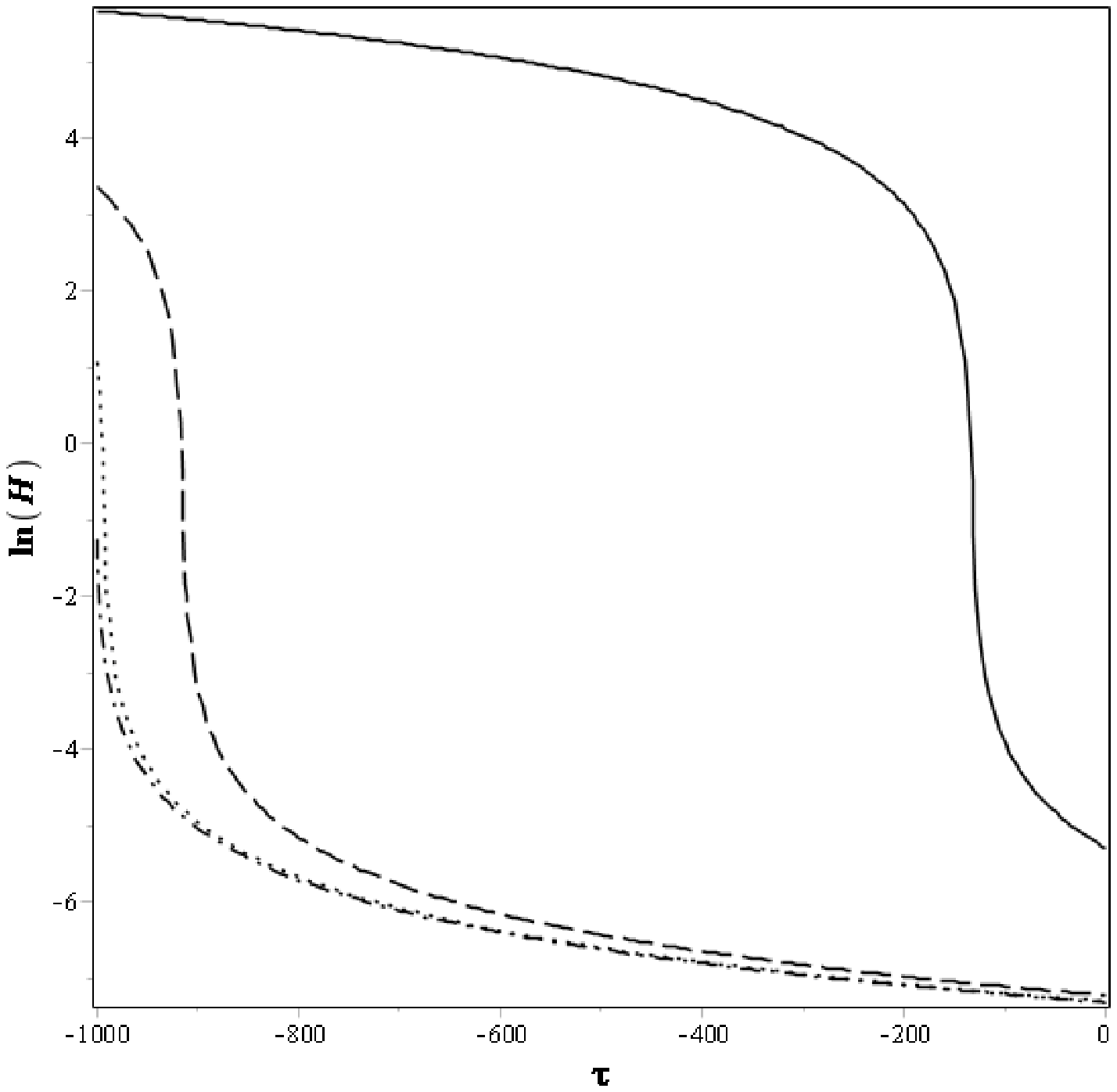}{7}{\label{ris23}Large-scale behavior of the logarithm of the normalized Hubble constant $h(\tau)$ at initial conditions: $\Phi(-1000)=100,\dot{\Phi}(-1000)=0$ is a solid line; $\Phi(-1000)=10,\dot{\Phi}(-1000)=0$ is a dashed line;$\Phi(-1000)=1,\dot{\Phi}(-1000)=0$ is a dotted line; $\Phi(-1000)=0.1,\dot{\Phi}(-1000)=0$ is a dotted-dashed line.}
Plots on Fig. \ref{ris23} also confirm the fact that the Universe's geometry tends asymptotically to Minkovsky geometry since it is
$H\to0\Rightarrow \dot(a)\to 0 \Rightarrow a\to \mathrm{Const}$.

In consequence of invariance of the cosmological acceleration $\Omega$ the cosmological acceleration can be calculated relative to scaling of time variable $t=\tau/m$ using formula coinciding with (\ref{Omega}), making the following substitution in it $H(t)\to h(\tau)$:
\begin{equation}\label{Omega_h}
\Omega=1+\frac{h'}{h^2}.
\end{equation}
Fig. \ref{ris25} shows a detailed behavior of the cosmological acceleration.

\Fig{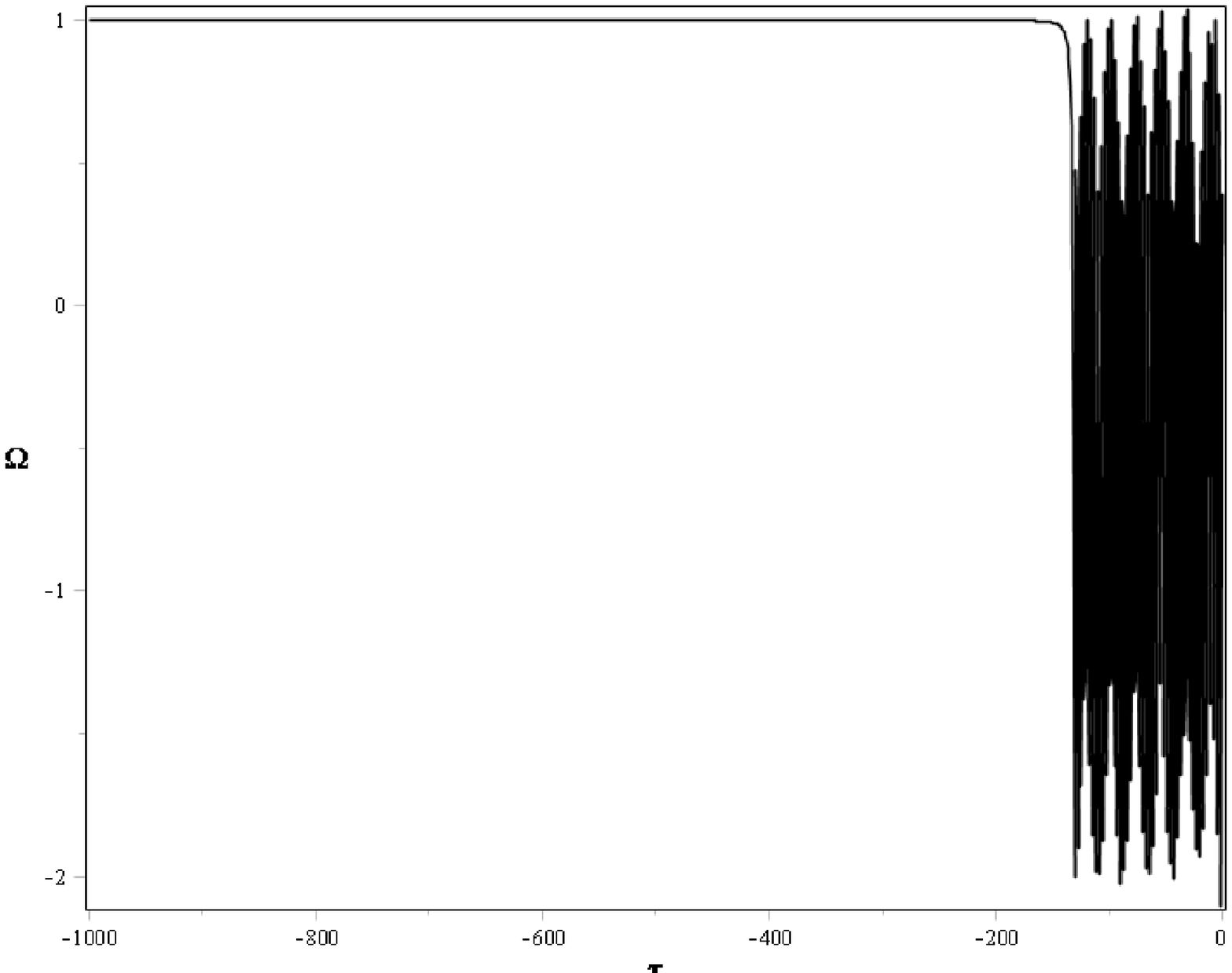}{7}{\label{ris24}Large-scale behavior of the cosmological acceleration $\Omega(\tau)$ at initial conditions $\Phi(-1000)=100,\dot{\Phi}(-1000)=0$.}
\Fig{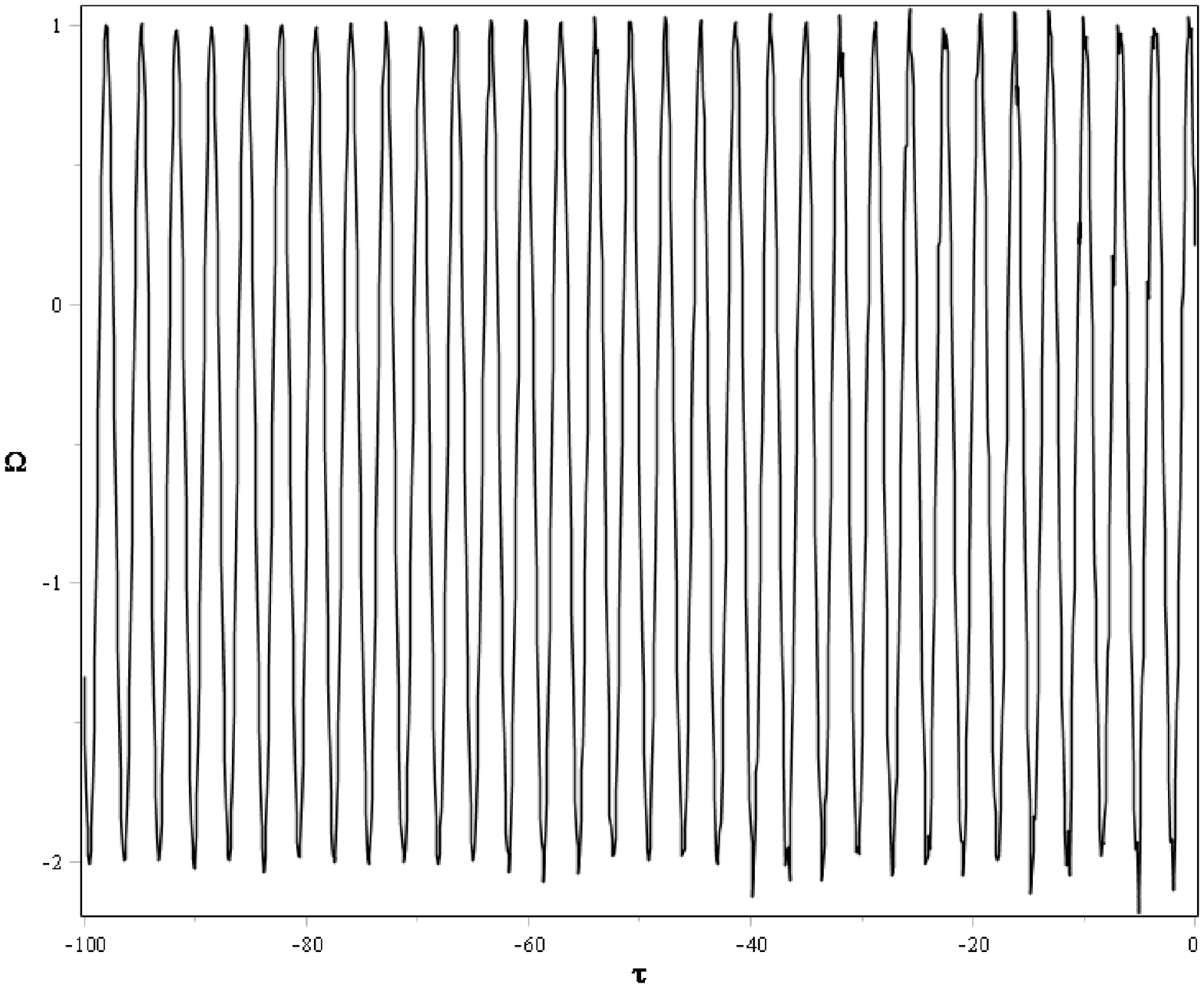}{7}{\label{ris25}Detailed behavior of the cosmological acceleration $\Omega(\tau)$ at the stage of oscillations at initial conditions $\Phi(-1000)=100,\dot{\Phi}(-1000)=0$.}
Herewith it needs to be understood that significant oscillations of the invariant cosmological acceleration at $\tau\to+\infty$ happen on the background of vanishingly small velocity of extension $H(+\infty)\to0$, therefore absolute values of the velocities tend to zero.

Further, the scale factor is obtained by integration (\ref{h}):
\[L(\tau)\equiv \ln a(\tau) =\int hd\tau.\]

\Fig{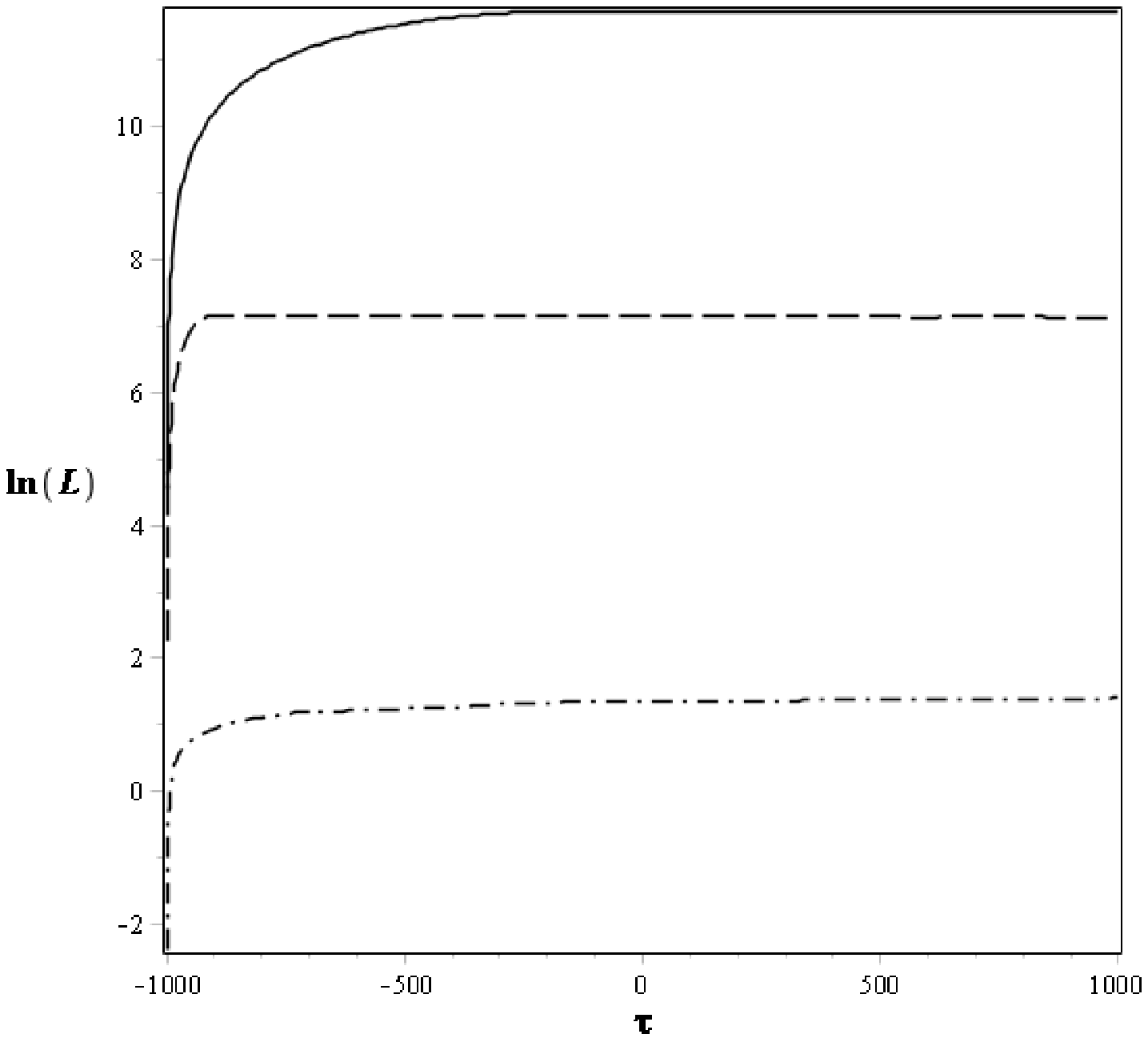}{7}{\label{ris26}The evolution of the scale factor at initial conditions: $\Phi(-1000)=100,\dot{\Phi}(-1000)=0$ -- ñïëîøíàÿ ëèíèÿ; $\Phi(-1000)=10,\dot{\Phi}(-1000)=0$ is a dashed line; $\Phi(-1000)=0.1,\dot{\Phi}(-1000)=0$ is a dotted-dashed line. The plot shows the values $\ln L$, where $L=\ln a(\tau)$.}
\subsection{The Average of the Cosmological Acceleration}
The fact that the invariant cosmological acceleration has an oscillating character at great times (Fig. \ref{ris25}), having the period of these oscillation on the time scale $\tau$ of the order of $2\pi$, i.e. in the ordinary time scale $T\sim2\pi/m$ being an obviously microscopic value, leads to necessity of introduction of the average value of the invariant cosmological acceleration averaged by large enough number of oscillations i.e. by large enough interval $\Delta\tau=N\cdot2\pi$, where $N \gg1$:
\begin{equation}\label{Omega_average0}
\overline{\Omega(\tau,\Delta\tau)}\equiv \frac{1}{\Delta\tau}\int\limits_{\tau}^{\tau+\Delta\tau}\Omega(\tau')d\tau'.
\end{equation}
Using formula (\ref{Omega_h}) in (\ref{Omega_average0}) and carrying out elementary integration, let us find for he average cosmological acceleration the following expression:
\begin{equation}\label{Omega_average}
\overline{\Omega(\tau,\Delta\tau)}=1+\frac{1}{\Delta\tau}\biggl(\frac{1}{h(\tau)}-\frac{1}{h(\tau+\Delta\tau)}\biggr).
\end{equation}
Fig. \ref{ris27} represents the plots of the average acceleration's dependency on time $\tau$.
\Fig{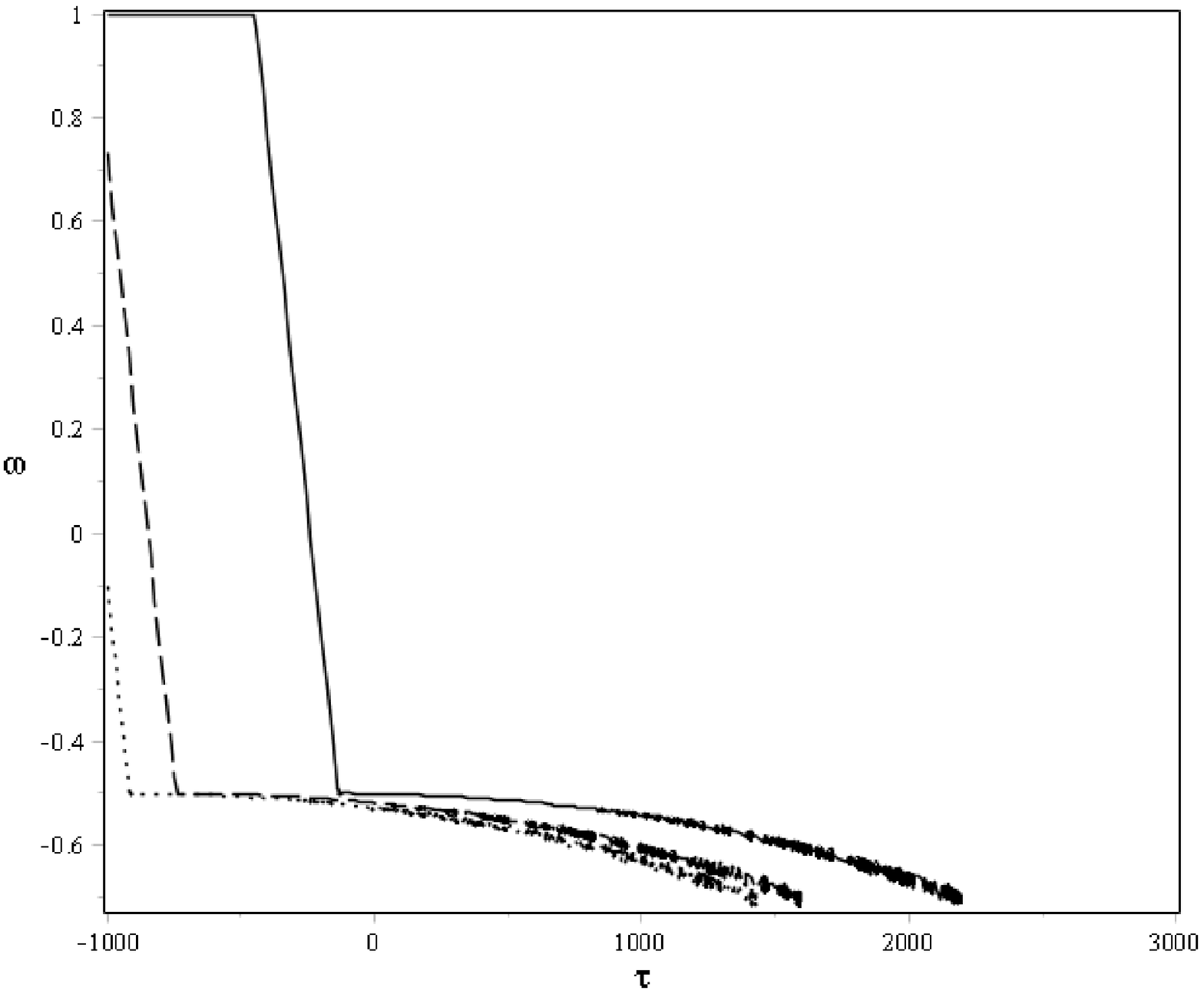}{7}{\label{ris27}Dependency of the average acceleration $\omega\equiv\overline{\Omega(\tau,\Delta\tau)}$ (\ref{Omega_average0}) on time $\tau$
at initial conditions: $\Phi(-1000)=100,\dot{\Phi}(-1000)=0$ is a solid line; $\Phi(-1000)=30,\dot{\Phi}(-1000)=0$ is a dashed line; $\Phi(-1000)=10,\dot{\Phi}(-1000)=0$ is a dotted line. It is everywhere accepted that $N=50$.}

\section{The Conclusion}
What can we conclude, looking at the last figure? First of all, we see that the macroscopic cosmological acceleration breaks down on finishing the inflation stage ($\overline{\Omega}=+1$) to value $\overline{\Omega}=-\frac{1}{2}$, and then slowly tends to $\overline{\Omega}=-1$. Secondly, let us notice that the invariant cosmological acceleration is connected with the barotropic coefficient $\kappa$ by means of the relation (\ref{Omega-kappa}) where from it follows that the value $\Omega=1$ corresponds to value $\kappa=-1$, value $\Omega=-1/2$ -- to value $\kappa=0$ and
$\Omega=-1$ -- to value $\kappa=1/3$.
\begin{thm}
Thus, the cosmological dynamic system founded on the Einstein - Klein - Gordon equations possesses the following property: the macroscopic evolution of the system consist from 3 clearly defined stages:
\item[1.] the inflation stage: $\overline{\Omega}=+1$; $\overline{\kappa}=-1$;
\item[2.] sharp transition to the non-relativistic stage: $\overline{\Omega}=-\frac{1}{2}$; $\overline{\kappa}=0$;
\item[3.] gradual slide seemingly to the ultrarelativistic stage: $\overline{\Omega}\to -1$; $\overline{\kappa}\to\frac{1}{3}$,
the Universe at that becomes asymptotically flat ($H(+\infty)\to 0$, $a(+\infty)\to \mathrm{Const}$).
\end{thm}
\section{Acknowledgments}
This work was funded by the subsidy allocated to Kazan Federal University for the state assignment in the sphere of scientific activities.

In conclusion, the Authors express their gratitude to the members of MW seminar for relativistic kinetics and cosmology of Kazan Federal University for helpful discussion of the work.


\end{document}